\documentclass[journal]{IEEEtran}

\usepackage{amsmath}    
\usepackage[pdftex]{graphicx}                
\usepackage{amssymb}                         
\usepackage{booktabs}                        
\usepackage{bm}                              
\usepackage{color}
\usepackage{enumerate}                       
\usepackage{epstopdf}                        
\usepackage{mathtools}                       
\usepackage[ruled,norelsize]{algorithm2e}    
\ifCLASSOPTIONcompsoc            
  \usepackage[caption=false,font=normalsize,labelfont=sf,textfont=sf]{subfig}
\else
  \usepackage[caption=false,font=footnotesize]{subfig}
\fi

\makeatletter
\newcommand{\removelatexerror}{\let\@latex@error\@gobble}
\makeatother

\begin{document}

\title{\huge MetaLocalization: Reconfigurable Intelligent Surface Aided Multi-user Wireless Indoor Localization}

\author{Haobo~Zhang,~\IEEEmembership{Student Member,~IEEE,}
        Hongliang~Zhang,~\IEEEmembership{Member,~IEEE,}
        Boya~Di,~\IEEEmembership{Member,~IEEE,}\\
        Kaigui~Bian,~\IEEEmembership{Senior Member,~IEEE,}
        Zhu~Han,~\IEEEmembership{Fellow,~IEEE,}
        and~Lingyang~Song,~\IEEEmembership{Fellow,~IEEE}
\thanks{Manuscript received August 27, 2020; revised January 14, 2021, April 4, 2021, and May 9, 2021; accepted June 2, 2021. This work was supported in part by the National Natural Science Foundation of China under Grants 61625101, 61931019, 61941101, and 62032003, and in part by NSF EARS-1839818, CNS1717454, CNS-1731424, and CNS-1702850. The associate editor coordinating the review of this paper and approving it for publication was Shaodan Ma. \emph{(Corresponding author: Lingyang Song.)}}
\thanks{H. Zhang, B. Di, and L. Song are with Department of Electronics, Peking University, Beijing, China (e-mail: haobo.zhang@pku.edu.cn, diboya92@gmail.com, lingyang.song@pku.edu.cn).}%
\thanks{H. Zhang is with Department of Electrical Engineering, Princeton University, NJ, USA (e-mail: hongliang.zhang92@gmail.com).}%
\thanks{K. Bian is with Department of Computer Science, Peking University, Beijing, China (e-mail: bkg@pku.edu.cn).}%
\thanks{Z. Han is with the Department of Electrical and Computer Engineering, University of Houston, Houston, TX 77004, USA, and also with the Department of Computer Science and Engineering, Kyung Hee University, Seoul, South Korea, 446-701 (e-mail: zhan2@uh.edu).}
}

\maketitle

\vspace{-6mm}

\begin{abstract}

  The received signal strength~(RSS) based technique is extensively utilized for localization in the indoor environments. 
  Since the RSS values of neighboring locations may be similar, the localization accuracy of the RSS based technique is limited. 
  To tackle this problem, in this paper, we propose to utilize reconfigurable intelligent surface~(RIS) for the RSS based multi-user localization. 
  As the RIS is able to customize the radio channels by adjusting the phase shifts of the signals reflected at the surface, the localization accuracy in the RIS aided scheme can be improved by choosing the proper phase shifts with significant differences of RSS values among adjacent locations.
  However, it is challenging to select the optimal phase shifts because the decision function for location estimation and the phase shifts are coupled. 
  To tackle this challenge, we formulate the optimization problem for the RIS-aided localization, derive the optimal decision function, and design the phase shift optimization (PSO) algorithm to solve the formulated problem efficiently. 
  Analysis of the proposed RIS aided technique is provided, and the effectiveness is validated through simulation.

\end{abstract}

\begin{IEEEkeywords}
Indoor localization, reconfigurable intelligent surface, received signal strength.
\end{IEEEkeywords}

%
\IEEEpeerreviewmaketitle

\vspace{-3mm}
\section{Introduction}
\label{s_introduction}

Future 6G wireless systems will be highly intelligent and support a wide range of applications such as human-centric mobile communications~\cite{Dang2020what}, personal navigation~\cite{Zafari2017A}, and healthcare monitoring~\cite{li2019convolutional}. 
To this end, the 6G wireless system has to not only realize enhanced ubiquitous communications but also provide accurate and reliable localization service, resulting in an increasing interest on wireless based localization techniques~\cite{Bourdoux20206G}.
Based on the utilized information for localization, these techniques can be categorized into several types such as the angle of arrival (AoA), time of arrival (ToA), or received signal strength (RSS) based techniques.
Since the RSS information can be easily extracted from the widespread Wi-Fi compatible devices without additional hardware requirements, the RSS based localization method is the widely used nowadays and is also expected to play an important role in future 6G wireless systems~\cite{yassin2017recent}.

\textcolor{black}{In literature, different types of RSS based wireless localization methods have been discussed~\cite{bahl2000radar}-\cite{qian2019multitarget}. 
For example, authors in~\cite{bahl2000radar} proposed a deterministic method that exploited the RSS information of the nearest neighborhood to estimate the location. 
The Bayesian network approach and the stored RSS distributions were adopted in~\cite{castro2001a} to infer the user's location. 
In~\cite{welss2003on}, authors considered the localization in a cellular system based on RSS measurements, and the localization accuracy was estimated by analyzing the Carmer-Rao bound, the concentration ellipse, and the circular error probability.
The authors in~\cite{ouyang2010received} studied the noncooperative and cooperative localization for a large number of wireless sensor nodes, and convex RSS estimators were utilized for localization.
In~\cite{coluccia2014rss}, a novel approach based on the empirical bayes and iterative least squares were proposed to improve the accuracy of RSS-based localization.
The authors in~\cite{qian2018compressive} and \cite{qian2019multitarget} studied the localization problem with imprecise location information of sensors, and utilized the compressive sensing method to effectively solve this problem. However, the localization performances in the aforementioned works are highly related to the RSS distribution. 
In the unfavorable RSS distribution where the RSS values of neighboring locations are similar, these locations become difficult to be distinguished, leading to the degradation of the localization accuracy.}

Fortunately, reconfigurable intelligent surfaces~(RISs) were recently developed to actively customize the radio environment~\cite{Shlezinger2019dynamic, Elmossallamy2020reconfigurable}, which become a promising solution to address this problem. 
To be specific, an RIS consists of a massive number of homogeneous elements in the form of a planar sheet~\cite{renzo2019smart}. 
By electrically tuning the state of each element, we can control its electromagnetic response (such as phase shifts), rendering different phases of the reflection signal given the incident signal~\cite{zhang2020reconfigurable}. 
Consequently, the RSS distribution in the surrounding radio environment can be customized, which shows the potential of the RIS to improve the localization accuracy of the RSS based technique by creating favorable RSS distributions.

\textcolor{black}{Several localization schemes aided by the RIS has been discussed in the literature. 
In~\cite{hu2018beyond}, large intelligent surfaces composed of active antenna elements were adopted to emit signals, and the received signal of the user was utilized for location estimation. 
The work~\cite{he2020large} investigated the RIS aided millimeter-wave (mmWave) multiple-input multiple-output (MIMO) positioning system, and the location was calculated based on the estimated channel gains, the angle of arrival (AoA), the angle of departure (AoD), and the time of arrival (ToA). 
The multiple-input single-output (MISO) counterpoint is considered in~\cite{fascista2020ris}.
The authors in~\cite{liu2020reconfigurable} proposed an RIS aided localization scheme which minimized the Carmer-Rao lower bound for high accuracy location estimation.
However, since the RSS information is not utilized in the aforementioned schemes, these schemes are not applicable for RSS based localization.}

In this paper, we consider an RIS aided multi-user localization scenario based on the RSS. 
An access point~(AP) emits signals which are reflected by the RIS to create favorable RSS distribution. 
Users can locate themselves by analyzing the measured RSS values, or cooperate with the RIS and the AP to iteratively improve the accuracy of the estimated locations. Compared with traditional RSS based techniques, the localization accuracy can be improved in the proposed scheme by carefully designing the RSS distributions with larger RSS differences.

However, several challenges need to be addressed for the proposed localization scheme. \emph{First}, different from traditional RSS based techniques without RIS, the operations of RIS in the proposed scheme are required to coordinate with those of AP and users to achieve high accuracy, which complicates the design of the RIS aided localization scheme.
\emph{Second}, the optimization of RIS phase shifts is challenging because the decision function for location estimation and the RSS distribution determined by the RIS phase shifts are coupled with each other.
\textcolor{black}{Besides, due to the discrete phase shifts of the RIS elements, the optimization of all the phase shifts is a non-linear integer program, which is NP-hard.}

To tackle these challenges efficiently, we design the RIS aided localization protocol, and formulate the optimization problem to minimize the weighted
probabilities of false localization, which is referred to as the localization loss.
To solve the formulated problem efficiently, we first derive the optimal decision function given phase shifts, and then find the optimal phase shifts utilizing the derived decision function and the proposed phase shift optimization~(PSO) algorithm. Overall, our contributions can be summarized below.
\begin{itemize}
  \item We propose to utilize the RIS for the RSS based multi-user localization in an indoor environment, and introduce a localization protocol to coordinate the operations of the AP, the RIS, and users during the localization process.
  \item We formulate the optimization problem for the multi-user localization, where the weighted probabilities of false localization is minimized by optimizing the decision function and the RIS phase shifts.
  \item \textcolor{black}{We derive the optimal decision function for location estimation, and propose the PSO algorithm to obtain the RIS phase shifts. Unlike most existing RIS optimization algorithms which are designed to maximize the sum-rate in communication systems, the PSO algorithm has different forms and is able to efficiently tackle the localization loss minimization problem.}
  \item We analyse the convergence, complexity, and the optimality of the PSO algorithm, and discuss the localization performance of the proposed localization scheme. The effectiveness of the proposed scheme is also verified through simulations. 
\end{itemize}

The rest of this paper is organized as follows. In Section~\ref{s_system}, we introduce the localization scenario and present the models of the RIS and the RSS. 
In Section~\ref{s_protocol}, a localization protocol is proposed. 
We formulate a optimization problem for the RIS-aided multi-user localization in Section~\ref{s_problem}. 
To solve the formulated problem, the optimal decision function is derived, and the PSO algorithm is designed in Section~\ref{s_algorithm}.
The analysis of the proposed scheme is provided in Section~\ref{s_analysis}. 
In Section~\ref{s_simulation}, we present the simulation results and discussions. Finally, the conclusions are drawn in Section~\ref{s_conclusion}.

\vspace{-2mm}
\section{System Model}
\label{s_system}

In this section, we first present the scenario of RIS aided multi-user localization in Section~\ref{ss_sd}. Then, we introduce the RIS model in Section~\ref{ss_RIS} and the RSS model in Section~\ref{ss_RSS}.

\vspace{-3mm}
\subsection{Scenario Description}
\label{ss_sd}

\begin{figure}[!t]
  \centering
  \includegraphics[width=3.3in]{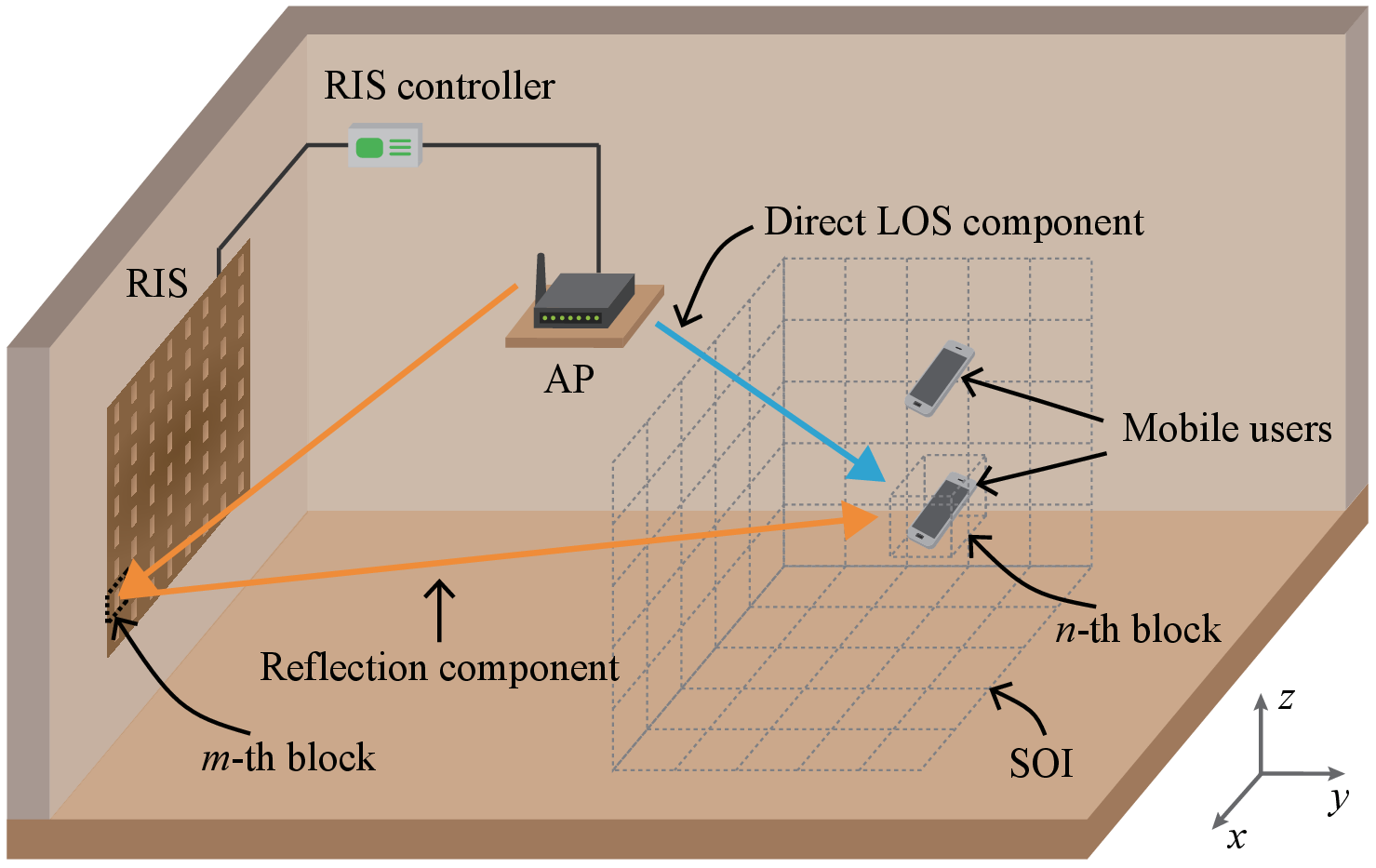}
  \caption{System model for the RIS aided multi-user localization.}
  \label{f_system}
\end{figure}

As shown in Fig.~\ref{f_system}, we consider a wireless indoor localization scenario which is composed of an AP, an RIS, and multiple users requiring their own indoor location information. The AP connects to the RIS controller which is able to regulate the operation of the RIS. During the localization process, the AP sends single-tone signal over frequency $f_c$ to the RIS and mobile users, and the RIS reflects the signals from the AP to the users. Each user measures the RSS for localization.

Specifically, all the mobile users are assumed to move in the space of interest~(SOI), which is a cubic region with size $l_x\times l_y\times l_z$. 
We discretize the SOI into $N$ blocks with the same size, and the set of blocks in the SOI is denoted by $\mathcal{N} = \{1, \cdots, N\}$. 
The location of each user can be represented by the index of the block where it is. 
By creating the favorable RSS distribution by the RIS, users at different blocks can be better distinguished comparing with traditional RSS based techniques.

\vspace{-3mm}
\subsection{RIS Model}
\label{ss_RIS}

The RIS is an artificial planar material consisting of a large number of elements made of metal and dielectric~\cite{Li2018Metasurfaces}, as illustrated in Fig.~\ref{f_system}. In each element, several subwavelength-scale metal patches are printed on the dielectric substrate which are connected by PIN diodes~\cite{tang2019wireless}. Each PIN diode can be electrically adjusted into two different states, i.e., the \emph{ON} and \emph{OFF} states~\cite{Li2019Machine}. The states of PIN diodes in an RIS element are referred to as the \emph{state} of the RIS element.  The signal reflected by the RIS element in different states has different phase shifts~\cite{zhang2021reconfigurable}.

Let $\mathcal{M} = \{1, \cdots, M\}$ denote the set of elements in the RIS, where $M$ is the number of RIS elements. We assume that each element has $C$ states with same amplitude ratio and uniform phase shift interval $\Delta\theta = \dfrac{2\pi}{C}$~\cite{di2019hybrid}. Therefore, the reflection coefficient of the $m$-th element can be expressed as
\begin{equation}
  r_m(c_m) = r e^{-jc_m\Delta\theta},
\end{equation}
where the amplitude ratio $r \in [0, 1]$ is a constant and $c_m \in \{1, \cdots, C\}$~\cite{hu2020reconfigurable}. For convenience, we use $c_m$ to represent the phase shift of the $m$-th element in this paper. Therefore, the vector of all the RIS elements' phase shifts can be expressed as $\bm{c} = (c_1, \cdots, c_m, \cdots, c_M)$.

\vspace{-3mm}
\subsection{RSS Model}
\label{ss_RSS}

In this subsection, we model the probability distribution of RSS in the SOI. As shown in Fig.~\ref{f_system}, the signal received by the user contains $M+1$ components: a direct line-of-sight (LOS) component and $M$ reflection components. The $m$-th reflection component is the transmission signal reflected off the $m$-th RIS element to the user. Since the transmitted signal is narrowband, the path loss \textcolor{black}{in dB} between the AP and the user at the $n$-th block can be expressed as\footnote{\textcolor{black}{In model (2), the first term denotes the mean path loss at the $n$-th block~\cite{goldsmith2005wireless}, and this term is derived using the analytical model~\cite{tang2020wireless}. Besides, the second term $\xi$ is the log-normal shadowing component which represents the multipath effects in the NLOS paths. Note that the path loss model (2) is different from the classical path loss model where the mean path loss is derived using the simplified path loss model.}}
\begin{equation}
  L_{n}(\bm{c}) \!=\! s_n(\bm{c}) \!-\! s^t \!=\! 20\log_{10}\left|h_{\mathrm{lo}} \!+\! \sum_{m\in\mathcal{M}} h_{m, n}(c_m)\right| \!+\! \xi,
  \label{e_received_signal}
\end{equation}
where $s^t$ is the transmitted signal power of AP, $s_n(\bm{c})$ is the RSS at the $n$-th block under phase shift vector $\bm{c}$, $h_{\mathrm{lo}}$ is the channel gain of the direct LOS component, $h_{m, n}(c_m)$ is the gain of the $m$-th reflection channel, and $\xi$ is the log-normal shadowing component which follows Gaussian distribution $\mathcal{N}(0, \sigma^2)$~\cite{Coluccia2018on}. For convenience, we define the mean RSS value at the $n$-th block under phase shift vector $\bm{c}$ as
\begin{equation}
  \mu_n(\bm{c}) \!=\! s^t \!+\! 20\log_{10}\left|h_{\mathrm{lo}} \!+\! \sum_{m\in\mathcal{M}} h_{m, n}(c_m)\right| \!=\! s_n(\bm{c}) \!-\! \xi.
  \label{def_mu}
\end{equation}

Based on \cite{goldsmith2005wireless}, the LOS channel gain $h_{\mathrm{lo}}$ can be expressed as
\begin{equation}
  h_{\mathrm{lo}} = \dfrac{\lambda}{4\pi} \cdot \dfrac{\sqrt{g^t_n g^r_n} \cdot e^{- j 2 \pi l_n / \lambda}}{l_n},\label{def_los_channel}
\end{equation}
where $\lambda$ is the wavelength of the transmitted signal, $g^t_n$ is the power gain of the AP antenna towards the $n$-th block, $g^r_{n}$ is the power gain of the user antenna at the $n$-th block towards the AP, and $l_n$ is distance between the AP and the user at the $n$-th block. Besides, the reflection channel gain $h_{m, n}(c_m)$ can be expressed as~\cite{basar2019wireless}
\begin{equation}
  h_{m, n}(c_m) \!=\! \dfrac{\lambda}{4\pi} \!\cdot\! \dfrac{\sqrt{g^t_m g^r_{m, n}} \!\cdot\! r_{m}(c_m) \!\cdot\! e^{- j 2 \pi (l^r_{m}+l^r_{m, n}) / \lambda}}{l^r_{m}l^r_{m, n}},
  \label{equ_mainChannelGain}
\end{equation}
where $g^t_m$ is the power gain of the AP antenna towards the $m$-th RIS element, $g^r_{m, n}$ is the power gain of the user antenna at the $n$-th block towards the $m$-th RIS element, $r_{m}(c_m)$ is the reflection coefficient of the $m$-th element in the state $c_m$, $l^r_m$ is the distance between the AP and the $m$-th RIS element, and $l^r_{m, n}$ is the distance between the $m$-th RIS element and the user at the $n$-th block.

Consequently, the probability distribution of RSS value at the $n$-th block under phase shift vector $\bm{c}$ can be expressed as
\begin{equation}
  \mathbb{P}(s_n(\bm{c}) = s) = \mathbb{P}(s | \bm{c}, n) = \dfrac{1}{\sqrt{2\pi\sigma^2}}e^{-\dfrac{(s-\mu_n(\bm{c}))^2}{2\sigma^2}},
  \label{e_rss_distribution}
\end{equation}
where $\sigma$ is the standard deviation of the RSS.

\section{RIS-aided Multi-user Localization Protocol}
\label{s_protocol}

\begin{figure}[!t]
  \centering
  \includegraphics[width=3.3in]{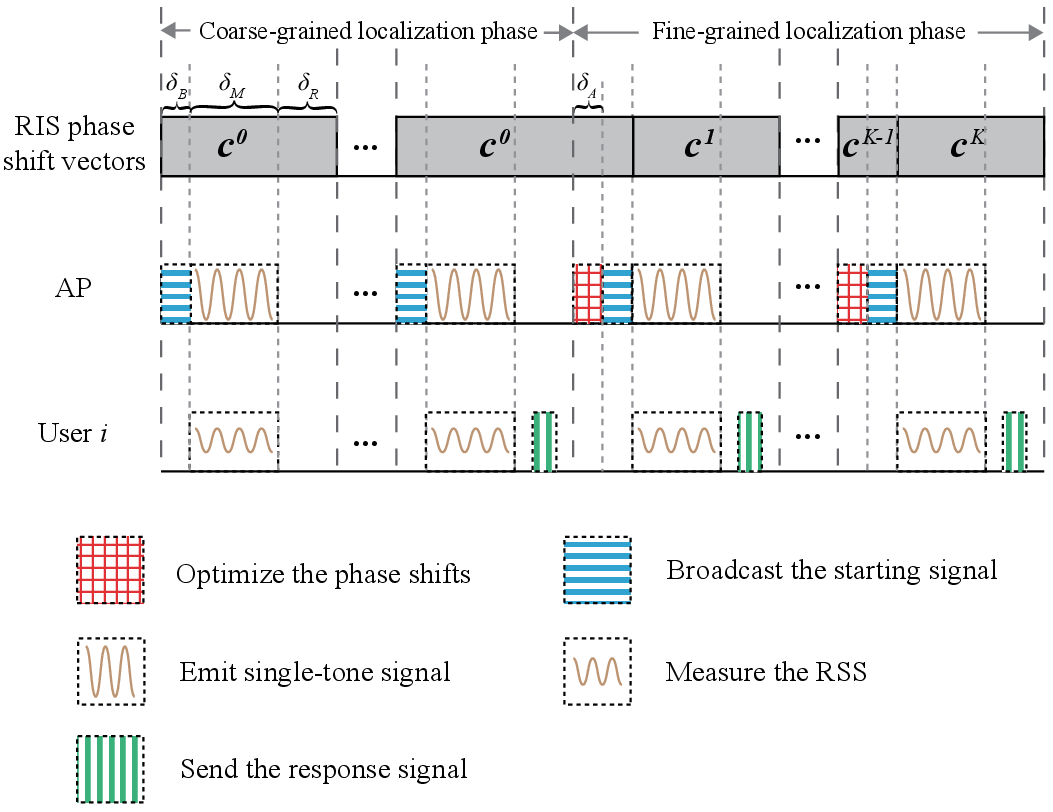}
  \caption{The RIS-aided multi-user localization protocol.}
  \label{f_sequence}
\end{figure}

In this section, we propose an RIS-aided multi-user localization protocol consisting of two operating phases: the \emph{coarse-grained} and the \emph{fine-grained localization phases}. In the coarse-grained localization phase in Section~\ref{ss_clp}, each user can locate itself using its measured RSS values within a short period of time. If any user requires location information with a higher precision, it will send a localization request to the AP, and the system will convert to the fine-grained localization phase, which is introduced in Section~\ref{ss_flp}. This phase relies on the cooperation among the AP, the RIS, and multiple users, and each user will be informed of its estimated location by the AP when the fine-grained localization phase terminates. The process of the RIS-aided multi-user localization protocol is illustrated in Fig.~\ref{f_sequence}.

\vspace{-3mm}
\subsection{Coarse-grained Localization Phase}
\label{ss_clp}

In this phase, the phase shift vector is fixed, which is denoted by $\bm{c}^0$. Based on ($\ref{e_received_signal}$)-($\ref{equ_mainChannelGain}$), the phase shift vector $\bm{c}^0$ can be utilized to calculate the mean RSS values in the SOI, which is referred to as the \emph{radio map}. Let $\bm{\mu}^0 = \{\mu_1(\bm{c}^0), \cdots, \mu_N(\bm{c}^0)\}$ denote the radio map under phase shift vector $\bm{c}^0$.
The timeline in this phase is divided into cycles, and users can conduct self-localization in every cycle. Three steps are conducted sequentially in each cycle for the coarse localization of multiple users.

\begin{enumerate}
  \item \textcolor{black}{\emph{Broadcast:} In the first $\delta_B$ seconds in each cycle, the AP broadcasts the starting signal to all users. The starting signals contains the information of the phase shift vector which can be used by the users to derive the radio map and locate themselves. Besides, The starting signal is also utilized to synchronize the AP and users.}  
  \item \emph{Measurement:} Then, in the next $\delta_M$ seconds, the AP sends single-tone signal with frequency $f_c$, and each user records the RSS during this period of time and computes the average RSS\footnote{\textcolor{black}{The duration of the measurement step $\delta_M$ needs to be less than the coherent time $T_D$ since the channel decorrelates after about $T_D$ seconds. According to~\cite{goldsmith2005wireless}, the coherent time of the channel $T_D \approx 0.4\lambda/v$, where $\lambda$ is the wavelength of the single-tone signal and $v$ is the speed of the user.}}. Let $s^0_i$ denote the average RSS for user $i$. With the average RSS $s^0_i$ and the radio map $\bm{\mu}^0$, user $i$ can estimate its location using the decision function $\mathcal{L}$, which will be introduced in the next section.
  \item \emph{Response:} In the last $\delta_R$ seconds in this cycle, if any user requires for location information with a higher precision, it will send out a response signal to the AP which contains the average RSS value in this cycle, as illustrated in Fig.~\ref{f_sequence}. To support multi-user communications, the time division multiplex (TDM) technique is adopted. Specifically, in the last $\delta_C - \delta_B - \delta_M$ seconds in every cycle, each user is assigned a time slot which is not overlapped with the time slots of other users. It is required to send the response signal during the assigned time slot. After receiving the response signal, the system will convert to the fine-grained localization phase.
\end{enumerate}

\vspace{-3mm}
\subsection{Fine-grained Localization Phase}
\label{ss_flp}

We still divide the timeline into cycles in this phase. The fine-grained localization phase will automatically terminate after $K$ cycles, and then the fine-grained localization results will be sent to users.

Different from the process in the coarse-grained localization phase, each cycle in the fine-grained localization phase contains four steps. Specifically, the optimization step is executed before the broadcast step, where the radio map and the phase shift vector in this cycle are carefully selected in order to promote the localization accuracy. Details of the four steps in this phase are introduced as follows.

\begin{enumerate}
  \item \emph{Optimization:} At the beginning of each cycle, AP selects the optimal phase shift vector for this cycle based on the RSS information collected in previous cycles. For the first cycle in the fine-grained localization phase, we use the RSS information collected in the coarse-grained localization phase. The optimization step lasts for $\delta_A$ seconds. Let $\bm{c}^k$ and $\bm{\mu}^k$ denote the phase shift vector and the corresponding radio map in the $k$-th cycle, respectively.
  \item \emph{Broadcast:} In the following $\delta_B$ seconds, the optimized phase shift vector $\bm{c}^k$ is sent to all the users and the RIS controller. The RIS controller will change the phase shifts of the RIS accordingly.
  \item \emph{Measurement:} For the next $\delta_M$ seconds, users will record the RSS while the AP emits signals with frequency $f_c$, which is similar to the measurement step in the coarse-grained localization phase. The average RSS of user $i$ in the $k$-th cycle is denoted by $s^k_i$.
  \item \emph{Response:} In this step, users are required to send the information of the average RSS to the AP in the assigned time slots. These average RSS values received by the AP will be utilized for the optimization in the next cycle.
\end{enumerate}

\vspace{-3mm}
\section{Problem Formulation}
\label{s_problem}

\newtheorem{proposition}{\bf Proposition}

In this section, we formulate the optimization problems for both the coarse-grained and the fine-grained localization phases, where the decision function and the RIS phase shift vector are optimized to improve the localization accuracy.

\vspace{-3mm}
\subsection{Problem Formulation for the Coarse-grained Localization Phase}

\textcolor{black}{To promote the localization accuracy in the coarse-grained localization phase, we minimize the \emph{localization loss} caused by false localization. Specifically, the localization loss is defined as the sum of expected localization error of all the users in one cycle, which can be expressed as
\begin{align}
  l(\bm{c}^0, \mathcal{L}) =~& \sum_{i \in \mathcal{I}}\sum_{\substack{n,n' \in \mathcal{N}\\ n\ne n'}} p^0_{i, n} \gamma_{n, n'} \notag\\
  & \times \int \mathbb{P}(s^0_i | \bm{c}^0, n) \mathcal{L}(n' | \bm{c}^0, s^0_i, \{p^0_{i, n}\})\cdot ds^0_i,
  \label{def_loss}
\end{align}
where $\mathcal{I} = \{1, \cdots, I\}$ denotes the set of all the mobile users, $p^0_{i, n}$ is the prior probability that user $i$ is located at the $n$-th block, and $\gamma_{n, n'}$ is the loss parameter when the estimated location is the $n'$-th block while the user is actually at the $n$-th block. The integral in (\ref{def_loss}) denotes the probability of false localization when we estimate the $n'$-th block as the location of user $i$ given that the ground truth of user $i$'s location is the $n$-th block. The decision function $\mathcal{L}(n' | \bm{c}^0, s^0_i, \{p^0_{i, n}\})$ estimates whether the location of user $i$ is the $n'$-th block given phase shift vector $\bm{c}^0$, RSS $s^0_i$, and prior probabilities $\{p^0_{i, n}\}$. Specifically, we have
\begin{equation}
  \mathcal{L}(n' | \bm{c}^0, s^0_i, \{p^0_{i, n}\}) = 
  \begin{cases}
    1, & \text{the estimated location of}\\
    & \text{user $i$ is the $n'$-th block,}\\
    0, & \text{otherwise.}
  \end{cases}
\end{equation}}

\textcolor{black}{Since there is no prior knowledge about the locations of users, we assume that users are uniformly distributed in the SOI. That is, $p^0_{i, n} = \dfrac{1}{N}, \forall i \in \mathcal{I}, \forall n \in \mathcal{N}$. The loss parameter is defined as the distance between the correct block and the misjudged block, which can be expressed as
\begin{equation}
  \gamma_{n, n'} = || \bm{r}_n - \bm{r}_{n'} ||,\label{def_gamma}
\end{equation}
where $\bm{r}_n$ is the location of the $n$-th block's center, and $|| \cdot ||$ denotes the Euclidean distance.}

Therefore, the optimization problem for the coarse-grained localization phase can be formulated as
\begin{subequations}
\begin{align}
  \text{(P1):}\min_{\bm{c}^0, \mathcal{L}}~& l(\bm{c}^0, \mathcal{L}),\label{p1_obj}\\
  s.t.~& \mathcal{L}(n' | \bm{c}^0,\! s^0_i,\! \{p^0_{i, n}\}) \!\in\! \{0,\! 1\},\!\forall i\!\in\! \mathcal{I},\! n,\! n'\!\in\!\mathcal{N},\label{p1_c1}\\
  & \sum_{n'\in\mathcal{N}}\!\mathcal{L}(n' | \bm{c}^0,\! s^0_i,\! \{p^0_{i, n}\}) = 1,\!\forall i\!\in\! \mathcal{I},\! n\!\in\!\mathcal{N},\label{p1_c2}\\
  & c^0_{m} \in \{1, \cdots, C\}, \forall m \in \mathcal{M},\label{p1_c3}\\
  & \gamma_{n, n'} = || \bm{r}_n - \bm{r}_{n'} ||, \forall n, n'\in\mathcal{N}.\label{p1_c4}
\end{align}
\end{subequations}

Here, constraints (\ref{p1_c1}) and (\ref{p1_c2}) are the properties of the decision function, constraint (\ref{p1_c3}) restricts the available states of the RIS elements, and constraint (\ref{p1_c4}) corresponds to the definition in (\ref{def_gamma}).

\vspace{-3mm}
\subsection{Problem Formulation for the Fine-grained Localization Phase}

As mentioned in Section~\ref{ss_flp}, in the fine-grained localization phase, the optimization needs to be conducted in every cycle based on the information collected in the previous cycles. In the $k$-th cycle, the optimization problem can be formulated as
\begin{subequations}
\begin{align}
  \text{(P2):}\min_{\bm{c}^k, \mathcal{L}}~& l(\bm{c}^k, \mathcal{L}) = \sum_{i \in \mathcal{I}}\sum_{\substack{n,n' \in \mathcal{N}\\ n\ne n'}} p^k_{i, n} \gamma_{n, n'} \notag\\
  &\times \int \mathbb{P}(s^k_i | \bm{c}^k, n) \mathcal{L}(n' | \bm{c}^k, s^k_i, \{p^k_{i, n}\})\cdot ds,\\
  s.t.~& \mathcal{L}(n' | \bm{c}^k,\! s^k_i,\! \{p^k_{i, n}\}\!)\! \in\! \{0,\! 1\},\!\forall i\!\in\! \mathcal{I},\! n,\! n'\!\in\!\mathcal{N},\label{p2_c1}\\
  & \sum_{n'\in\mathcal{N}}\!\mathcal{L}(n' | \bm{c}^k,\! s^k_i,\! \{p^k_{i, n}\}) = 1,\!\forall i\!\in\! \mathcal{I},\! n\!\in\!\mathcal{N},\label{p2_c2}\\
  & c^k_{m} \in \{1, \cdots, C\}, \forall m \in \mathcal{M},\label{p2_c3}\\
  & \gamma_{n, n'} = || \bm{r}_n - \bm{r}_{n'} ||,\forall n, n'\in\mathcal{N},\label{p2_c4}
\end{align}
\end{subequations}
where $p^{k}_{i, n}$ denote the prior probability in the $k$-th cycle, $\mathcal{L}(n' | \bm{c}^k, s^k_i, \{p^k_{i, n}\})$ is the decision function which estimates whether the location of user $i$ is the $n'$-th block in the $k$-th cycle, and the localization results in the fine-grained localization phase can be given by $\mathcal{L}(n' | \bm{c}^K, s^K_i, \{p^K_{i, n}\}), \forall n, n'$. Constraints (\ref{p2_c1})-(\ref{p2_c4}) are similar to those in (P1).

The prior probabilities in the $k$-th cycle imply our belief about the probability distribution of users' locations based on the radio maps and RSS values in the previous cycles. According to Bayes' theorem, the prior probability that user $i$ is at the $n$-th block in the $k$-th cycle can be expressed as~\cite{papoulis2002probability}
\begin{equation}
  p^k_{i, n} \!\approx\! \mathbb{P}(n | \bm{c}^{k-1},\! s^{k-1}_i) \!=\! \dfrac{p^{k-1}_{i, n} \mathbb{P}(s^{k-1}_i | \bm{c}^{k-1}, n)}{\sum_{n\in\mathcal{N}} p^{k-1}_{i, n} \mathbb{P}(s^{k-1}_i | \bm{c}^{k-1}, n)}.
  \label{d_prior_probability}
\end{equation}
Here, $p^0_{i, n}$ is the prior probability in the coarse-grained localization phase.

\vspace{-3mm}
\section{Algorithm Design}
\label{s_algorithm}

In this section, we design the algorithm for the above-mentioned optimization problems. We can observe that (P1) is a special case of (P2), and thus we only elaborate on the algorithm to solve problem (P2) in the following. We first optimize the decision function in Section~\ref{ss_df}, and then propose the phase shift optimization (PSO) algorithm to select the most suitable phase shift vector in Section~\ref{ss_psoa}. For simplicity, we omit the superscript $k$ for cycles in this section.

\vspace{-3mm}
\subsection{Decision Function}
\label{ss_df}

Given phase shift vector $\bm{c}$, RSS $s_i$ and prior probabilities $\{p_{i, n}\}$, the optimal decision function $\mathcal{L}^*(n' | \bm{c}, s_i, \{p_{i, n}\})$ for problem (P2) is given by the following proposition:

\begin{proposition}
  \label{pro_opt_decision}
  The optimal decision function $\mathcal{L}^*(n' | \bm{c}, s_i, \{p_{i, n}\})$ for problem (P2) can be expressed as
  \begin{equation}
    \mathcal{L}^*(n' | \bm{c}, s_i, \{p_{i, n}\}) =
    \begin{cases}
      1, & s_i \in \mathcal{R}_{i, n'},\\
      0, & s_i \notin \mathcal{R}_{i, n'},
    \end{cases}
  \end{equation}
  where the decision region $\mathcal{R}_{i, n'}$ is defined as\textcolor{black}{
  \begin{align}
    \mathcal{R}_{i, n'} = \bigg\{& s_i : \sum_{n\in\mathcal{N}} p_{i, n} (\gamma_{n,n'} - \gamma_{n,n''}) \mathbb{P} (s_i | \bm{c}, n) \le 0, \notag\\
    &\forall n'' \in \mathcal{N}/\{n'\}\bigg\}.\label{pro_opt_decision_c1}
  \end{align}}
\end{proposition}
\begin{IEEEproof}
  See Appendix \ref{proof_opt_decision}.
\end{IEEEproof}

Therefore, the localization loss can be expressed as
\begin{equation}
  l(\bm{c}, \mathcal{L}^*) \!=\! \sum_{i \in \mathcal{I}}\!\sum_{\substack{n,n' \in \mathcal{N}\\ n\ne n'}} p_{i, n} \gamma_{n, n'} \int_{s_i \in \mathcal{R}_{i, n'}} \!\mathbb{P}(s_i | \bm{c}, n) \cdot ds_i.\label{def_loss2}
\end{equation}

However, since the decision region is irregular, the integration in (\ref{def_loss2}) is hard to tackle. An approximation of the localization loss is provided by the following proposition:
\begin{proposition}
  \label{pro_approx_loss}
  An approximation for the localization loss can be expressed as
  \begin{align}
    l_a(\bm{c}) &= \sum_{i \in \mathcal{I}} \sum_{\substack{n,n' \in \mathcal{N}\\ n\ne n'}} p_{i, n} \gamma_{n, n'} \cdot Q(d_{i, n, n'}) \ge l(\bm{c}, \mathcal{L}^*),
    \label{def_approx_loss}
  \end{align}
  where $Q(\cdot)$ is the Gaussian Q function~\cite{simon2004digital}, and $d_{i, n, n'}$ can be expressed as
  \begin{equation}
    d_{i, n, n'} = \dfrac{(\mu_{n'} - \mu_n)^2 - 2\sigma^2 \ln \dfrac{p_{i, n'}}{p_{i, n}}}{2\sigma|\mu_{n'}-\mu_n|}.
    \label{def_d}
  \end{equation}
\end{proposition}
\begin{IEEEproof}
  See Appendix \ref{proof_approx_loss}.
\end{IEEEproof}

In the following, we use the approximated localization loss $l_a$ to replace the original localization loss $l$ in order to solve the original problem more efficiently.

\vspace{-3mm}
\subsection{Phase Shift Optimization Algorithm}
\label{ss_psoa}

\textcolor{black}{Based on the approximated localization loss $l_a$, the phase shift optimization problem can be formulated as}
\begin{subequations}
  \begin{align}
    \text{(P3):}\min_{\bm{c}}~& l_a(\bm{c}),\\
    s.t.~&c^k_{m} \in \{1, \cdots, C\}, \forall m \in \mathcal{M},\label{p3_c1}\\
    & \gamma_{n, n'} = || \bm{r}_n - \bm{r}_{n'} ||,\forall n, n'\in\mathcal{N}.\label{p3_c2}
  \end{align}
\end{subequations}
Due to the non-convex objective function, problem (P3) can be proved to be NP-hard~\cite{arora2009computational}. Besides, problem (P3) is an integer optimization problem, which is more difficult to deal with compared to the continuous optimization problem~\cite{conforti2009integer}. 

To solve this problem efficiently, we propose a PSO algorithm based on the global descent method~\cite{ng2007discrete}. The PSO algorithm consists of two phases: initialization phase and global search phase. In the initialization phase, we use the local search method to generate a set of local minimum phase shift vector. The definition of the local minimum phase shift vector is provided as follows.

\newtheorem{definition}{\bf Definition}

\begin{definition}
  The vector $\bm{c}^*$ is referred to as a local minimum phase shift vector if 
  \begin{equation}
    l_a(\bm{c}^*) \le l_a(\bm{c}) + \epsilon, \forall \bm{c} \in \mathcal{U}(\bm{c}^*),
  \end{equation}
  where $\epsilon$ is a small but nonzero constant, and $\mathcal{U}(\bm{c}^*)$ is the unit neighborhood of phase shift vector $\bm{c}^*$, which can be expressed as
  \begin{equation}
    \mathcal{U}(\bm{c}^*) = \{\bm{c}\ |\ (\bm{c} - \bm{c}^*)\ mod\ C = \pm \bm{e}_m, m \in \mathcal{M}\},
  \end{equation}
  where \emph{mod} is the modulo operator, and the vector $\bm{e}_m$ is a unit vector. Specifically, the $m$-th element in $\bm{e}_m$ is $1$, and other elements in $\bm{e}_m$ is $0$.
\end{definition}

In the global search phase, the generated local minimum phase shift vectors will be utilized to conduct global search to approach the global minimum of the localization loss. 

\subsubsection{Initialization Phase}

\begin{figure}[!t]
  \removelatexerror
  \begin{algorithm}[H]
    \caption{Local Minimum Phase Shift Vector Search Algorithm}
    \label{a_lmcs}
    \KwIn{Inital phase shift vector $\bm{c}'$;}
    \KwOut{Local minimum phase shift vector $\bm{c}^*$;}
    Initial $z = 0$, $\bm{c}^0 = \bm{c}'$\;
    \While{$\bm{c}^z$ is not a local minimum phase shift vector}{
      Set $\bm{c}^{z+1}$ as the phase shift vector with the minimum loss in $\mathcal{U}(\bm{c}^z)$\;
      Update $z = z + 1$\;
    }
    Set $\bm{c}^* = \bm{c}^z$\;
  \end{algorithm}
\end{figure}

In this phase, we first use the local minimum phase shift vector search (LMVS) algorithm to obtain $Z^l$ different local minimum phase shift vectors. The set of $Z^l$ local minimum phase shift vectors is denoted by $\mathcal{C}$. The phase shift vectors in $\mathcal{C}$ are then sorted in an increasing order according to their localization loss.

The LMVS algorithm can find a local minimum phase shift vector from an initial phase shift vector input using the alternating optimization method~\cite{Bezdek2002Some}. Specifically, if the input phase shift vector $\bm{c}^0$ is not a local minimum phase shift vector, the algorithm will find a phase shift vector $\bm{c}^1$ which is the phase shift vector in $\mathcal{U}(\bm{c}^0)$ with the minimum positioning loss. If $\bm{c}^1$ is a local minimum phase shift vector, the LMVS algorithm terminates and the output phase shift vector is $\bm{c}^1$. Otherwise, the algorithm will search in the unit neighborhood $\mathcal{U}(\bm{c}^1)$ to find a new phase shift vector $\bm{c}^2$ and decide whether $\bm{c}^2$ is a local minimum phase shift vector. The procedures of the LMVS algorithm is summarized as Algorithm~\ref{a_lmcs}.

\begin{figure}[!t]
  \removelatexerror
  \begin{algorithm}[H]
    \caption{Phase Shift Optimization Algorithm}
    \label{a_co}
    \KwIn{Parameter $Z^u$;}
    \KwOut{Phase shift vector $\bm{c}^*$;}
    Initial a set of $Z^l$ different local minimum phase shift vectors denoted by $\mathcal{C}$ using the LMVS algorithm and random phase shift vector inputs.\;
    Sort phase shift vectors in $\mathcal{C}$ in increasing order of their localization loss\;
    \While{$|\mathcal{C}| \le Z^u$}{
      For the first phase shift vector $\bm{c}^f$ in $\mathcal{C}$, compute the descent ratios $r(\bm{c}^f, \bm{c}), \forall \bm{c} \in \mathcal{C}/\{\bm{c}^f\}$\;
      Sort the descent ratios in a descending order, and choose the phase shift vector $\bm{c}^m$ with the maximum descent ratio $r(\bm{c}^f, \bm{c}^m)$\;
      Calculate the steepest descent direction $\bm{d} = (\bm{c}^m - \bm{c}^f)\ mod\ C$\;
      Enumerate $\zeta \in \{1, \cdots, C\}$ to find phase shift vector $\bm{c}^s = (\bm{c}^f + \zeta\bm{d})~mod~C$ with the minimum localization loss, and calculate the local minimum phase shift vector $\bm{c}'$ using the LMVS algorithm with input $\bm{c}^s$\;
      \eIf{$\bm{c}' \notin \mathcal{C}$}{
        Insert $\bm{c}'$ into sorted set $\mathcal{C}$ according to $l_a(\bm{c}')$\;
      }{
        Generate a new local minimum phase shift vector $\bm{c}''$ which is not in $\mathcal{C}$ using the LMVS algorithm, and insert $\bm{c}''$ into the sorted set $\mathcal{C}$ according to $l_a(\bm{c}'')$\;
      }
    }
    Set $\bm{c}^*$ as the first phase shift vector in the set $\mathcal{C}$\;
  \end{algorithm}
\end{figure} 

\subsubsection{Global Search Phase}

In this phase, the algorithm will iteratively infer other local minimum phase shift vectors using the phase shift vectors in the set $\mathcal{C}$. The method to infer other local minimum phase shift vectors is inspired by the steepest descent method for the continuous optimization problems~\cite{boyd2004convex}. In each iteration, three steps are conducted sequentially.

\begin{itemize}
  \item The descent ratios between the first phase shift vector and other phase shift vectors in the set $\mathcal{C}$ is computed first. The descent ratio between phase shift vectors $\bm{c}^f$ and $\bm{c}$ is defined as
  \begin{equation}
    r(\bm{c}^f, \bm{c}) = \dfrac{l_a(\bm{c}) - l_a(\bm{c}^f)}{||\bm{c} - \bm{c}^f||},
  \end{equation}
  where $\bm{c}^f$ is the first phase shift vector in $\mathcal{C}$, and phase shift vector $\bm{c} \in \mathcal{C}/\{\bm{c}^f\}$. Let $\bm{c}^m$ denote the phase shift vector with the maximum descent ratio with $\bm{c}^f$. Consequently, the steepest descent direction can be expressed as $\bm{d} = (\bm{c}^m - \bm{c}^f)~mod~C$.
  \item Next, we enumerate the step size $\zeta \in \{1, \cdots, C\}$ to find a new phase shift vector $\bm{c}^s = ((\bm{c}^f + \zeta\bm{d})~mod~C)$ with the minimum localization loss. Using the LMVS algorithm, we can find a new local minimum phase shift vector $\bm{c}'$ with input $\bm{c}^s$.
  \item If $\bm{c}' \notin \mathcal{C}$, the phase shift vector $\bm{c}'$ will be inserted into the sorted set $\mathcal{C}$ according to the value of its localization loss $l_a(\bm{c}')$. Otherwise, we use the LMVS algorithm with random inputs to a new local minimum phase shift vector $\bm{c}''$ with is not in $\mathcal{C}$, and then phase shift vector $\bm{c}''$ will be inserted into the sorted set $\mathcal{C}$.
\end{itemize}
 
The iteration terminates when the number of phase shift vectors in the set $\mathcal{C}$ is greater than $Z^u$, and we have $Z^u > Z^l$. After the iteration ends, the algorithm will output the first phase shift vector in the set $\mathcal{C}$. The procedures of the PSO algorithm is summarized as Algorithm~\ref{a_co}.

\vspace{-3mm}
\section{Performance Analysis}
\label{s_analysis}

In this section we analyse the convergence, complexity, and optimality of the proposed algorithms, and discuss the localization performance of the proposed scheme.

\vspace{-3mm}
\subsection{Algorithm Convergence}

\subsubsection{Convergence of the LMVS Algorithm}

In the $z$-th iteration, a phase shift vector denoted by $\bm{c}^z$ with the minimal localization loss among $2M$ different phase shift vectors is obtained. If $l_a(\bm{c}^z) \le l_a(\bm{c}) + \epsilon, \forall \bm{c} \in \mathcal{U}(\bm{c}^z)$, a local minimum phase shift vector is found, and the algorithm converges. Otherwise, a phase shift vector $\bm{c}^{z+1}$ can be obtained in the $(z+1)$-th iteration with lower localization loss comparing to $\bm{c}^z$ because $l_a(\bm{c}^z) > l_a(\bm{c}) + \epsilon, \exists \bm{c} \in \mathcal{U}(\bm{c}^z)$. Therefore, we have $l_a(\bm{c}^{z+1}) + \epsilon < l_a(\bm{c}^z)$, which indicates that the localization loss decreases when the number of iteration increases. Since the localization loss is greater than $0$ for any phase shift vector, the algorithm is guaranteed to converge.

\subsubsection{Convergence of the PSO Algorithm}

Since the PSO algorithm terminates after $(Z^u - Z^l + 1)$ iterations, it will converge if each iteration of the PSO algorithm converges. In each iteration, the algorithm needs to find a new local minimum phase shift vector which is not in the set $\mathcal{C}$. Since the number of possible phase shift vectors $C^M \gg Z^u$, which implies that the number of local minimum phase shift vectors is far greater than $Z^u$, the algorithm can always obtain a new local minimum phase shift vector by randomly choosing initial phase shift vectors for the LMVS algorithm. Therefore, the convergence of the PSO algorithm is guaranteed.

\vspace{-3mm}
\subsection{Algorithm Complexity}

\subsubsection{Complexity of the LMVS Algorithm} Since the localization loss is reduced by at least $\epsilon$ in each iteration, the LMVS Algorithm has at most $l^u_a / \epsilon$ iterations, where $l^u_a$ is the upper bound of the localization loss. In Proposition~\ref{pro_upper}, an upper bound is provided for the localization.
\begin{proposition}
  \label{pro_upper}
  An upper bound of localization loss $l^u_a$ can be expressed as
  \begin{equation}
    l_a(\bm{c}) < l^u_a = I N \sqrt{l^2_x + l^2_y + l^2_z}, \forall \bm{c}.
  \end{equation}
\end{proposition}
\begin{IEEEproof}
  See Appendix \ref{proof_upper}.
\end{IEEEproof}

In each iteration of the LMVS Algorithm, the localization loss is calculated for $2M$ times. According to (\ref{def_approx_loss}), the complexity to calculate the localization loss for one phase shift vector is $O(IN^2)$. Therefore, the time complexity of each iteration is $O(IMN^2)$, and the time complexity of the LMVS algorithm is $O(I^2 M N^3)$.

\subsubsection{Complexity of the PSO Algorithm}
\label{ss_complexity_ro}

In the initialization phase, the LMVS algorithm is first conducted $Z^l$ times, and the complexity of this step is $O(Z^lI^2 M N^3)$. As for the phase shift vector sorting in the set $\mathcal{C}$, we assume its complexity is $O((Z^l)^2)$.

In the global search phase, there are $Z^u+1-Z^l$ iterations. In each iteration, the phase shift vector with maximum descent ratio is chosen. Since there are at most $Z^u$ phase shift vectors, the time complexity of this step is $O(Z^u)$. Next, the localization loss of $C-1$ phase shift vectors is calculated to determine the optimal step size, which has the time complexity $O(C I N^2)$. Note that we assume that the phase shift vector obtained by the LMVS algorithm with random initial phase shift vector is not in the set $\mathcal{C}$ because the number of possible phase shift vector $C^M \gg Z^u$. Thus, the LMVS algorithm is conducted at most twice to find a local minimum phase shift vector which can be inserted into the set $\mathcal{C}$, and the time complexity of this step is $O(I^2 M N^3)$. To sum up, the time complexity of the global search phase is $O((Z^u+1-Z^l)(I^2 M N^3))$.

\vspace{-3mm}
\subsection{Optimality}

In this subsection we analyse the optimality of the PSO algorithm. 

\begin{proposition}
  \label{pro_less}
  \textcolor{black}{In each iteration of the global search phase in the PSO algorithm, we have 
  \begin{equation}
    \mathbb{E}(l_a(\bm{c}')) < \mathbb{E}(l_a(\bm{c}^f)) < \mathbb{E}(l_a(\bm{c}'')),
  \end{equation}
  where $\mathbb{E}(\cdot)$ is the expectation operator, $\bm{c}'$ is the phase shift vector generated using the steepest descent method, $\bm{c}^f$ is the phase shift vector with the minimum localization loss in $\mathcal{C}$, and $\bm{c}''$ is the phase shift vector generated using the LMVS algorithm with random phase shift vector input.}
\end{proposition}
\begin{IEEEproof}
  See Appendix~\ref{proof_less}.
\end{IEEEproof}

\textcolor{black}{According to proposition~\ref{pro_less}, we can show that $\mathbb{E}(l_a(\bm{c}^{f, z}))$ decreases when $z$ increases. Specifically, let $\mathcal{C}^z$ and $\mathcal{C}^{z+1}$ denote the sorted set in the $z$-th and $(z+1)$-th iteration, respectively. Besides, let $\bm{c}^{f, z}$ and $\bm{c}^{f, z+1}$ denote the first vector in the sorted set $\mathcal{C}^z$ and $\mathcal{C}^{z+1}$, respectively. In the $z$-th iteration of the PSO algorithm, the vector $\bm{c}'$ or $\bm{c}''$ will be added in the sorted set $\mathcal{C}^z$ to generate $\mathcal{C}^{z+1}$. According to proposition~\ref{pro_less}, we have $\mathbb{E}(l_a(\bm{c}^{f, z+1})) < \mathbb{E}(l_a(\bm{c}^{f, z}))$, which indicates that when $z$ increases, $\mathbb{E}(l_a(\bm{c}^{f, z}))$ will decrease.}

\textcolor{black}{For each realization, we have $l_a(\bm{c}^{f, z+1}) \le l_a(\bm{c}^{f, z})$ based on the proof of proposition 4, which means that $l_a(\bm{c}^{f, z})$ is non-increasing when $z$ increases. However, since the expectation of $l_a(\bm{c}^{f, z})$ decreases when $z$ increases, $l_a(\bm{c}^{f, z})$ will decrease after sufficient number of iterations, and thus each realization will approach the global optimality when $Z^u$ is sufficiently large.}

\vspace{-3mm}
\subsection{Localization Performance}

In this subsection, we analyse the the localization performance in the coarse-grained and fine-grained localization phase. To evaluate the performance of multi-user localization, we define the localization error as
\begin{equation}
  l_e = \dfrac{1}{I}\sum_{i\in\mathcal{I}}||\bm{r}^e_i - \bm{r}^g_i||,
  \label{def_error}
\end{equation}
where $\bm{r}^e_i$ is the location of the estimated block's center for user $i$, and $\bm{r}^g_i$ is the ground truth.

\subsubsection{Localization Performance of the Coarse-grained Localization Phase}
\label{sss_coarse}

According to~(\ref{def_gamma}), (\ref{def_approx_loss}), and (\ref{def_error}), the expectation of the localization error is proportional to the localization loss, which can be expressed as
\begin{equation}
  \textcolor{black}{\mathbb{E}(l_e) \!=\! \dfrac{l(\bm{c},\! \mathcal{L}^*\!)}{I} \!\approx\! \dfrac{l_a(\bm{c})}{I} \!=\!\! \dfrac{1}{IN} \sum_{i \in \mathcal{I}}\! \sum_{\substack{n,n' \!\in \mathcal{N}\\ n\ne n'}}\!\! \gamma_{n, n'} \!\cdot\! Q(d_{i, n, n'}\!),}
  \label{exp_expect_error}
\end{equation}
where $d_{i, n, n'} = \dfrac{|\mu_n - \mu_{n'}|}{2\sigma}$ because $p_{i, n} = \dfrac{1}{N}, \forall n \in \mathcal{N}$. 

The expression of (\ref{exp_expect_error}) shows that the expectation of localization error $\mathbb{E}(l_e)$ is negatively related to the RSS differences $|\mu_n - \mu_{n'}|$. As a result, the performance of traditional RSS based schemes are degraded if the RSS differences are small. However, by integrating the RIS into the RSS based scheme, the RSS differences $|\mu_n - \mu_{n'}|$ can be enlarged to reduce the expectation of localization error $\mathbb{E}(l_e)$. 

Let $d^{min}$ denote the minimum $d_{i, n, n'}$ for all blocks. Therefore, an upper bound of the localization error can be expressed as
\begin{equation}
  \mathbb{E}(l_e) \le \dfrac{1}{IN} \sum_{i \in \mathcal{I}} \sum_{\substack{n,n' \in \mathcal{N}\\ n\ne n'}} \gamma_{n, n'} \cdot Q(d^{min}).
  \label{exp_upper_loss}
\end{equation}
Note that the value of mean RSS is limited. Let $\mu^{max}$ and $\mu^{min}$ denote the maximum and minimum possible mean RSS values, respectively. The corresponding maximum $d^{min}$ is 
\begin{equation}
  d^{min} = \dfrac{\mu^{max} - \mu^{min}}{2\sigma(N-1)}.
  \label{exp_d_min}
\end{equation}
The maximum mean RSS value $\mu^{max}$ is realized when the phases of the signals reflected by the RIS elements are aligned with the phase of the LOS signal, which can be expressed as
\begin{equation}
  \mu^{max} = s^t + 20\log_{10}\left||h_{\mathrm{lo}}| + \sum_{m\in\mathcal{M}} |h_{m, n^*}(c_m)|\right|,
  \label{def_mu_max}
\end{equation}
where $n^*$ denotes the block that maximizes (\ref{def_mu_max}). Since the $|h_{m, n^*}(c_m)|$ is positive given any $m$ and $n$, $\mu^{max}$ is positively related to the number of RIS elements. Besides, $\mu^{max}$ is negatively related to the distances among AP, RIS, and users according to (\ref{def_los_channel}) and (\ref{equ_mainChannelGain}). As for $\mu^{min}$, it can be achieved when the LOS signal and the reflection signals are cancelled, which rely on the precise adjustment of the phase shifts of the RIS elements.

Consequently, we have the following remark for the localization error in the coarse-grained localization phase:
\newtheorem{remark}{\bf Remark}
\begin{remark}
  \label{re_error_coarse}
  The localization error in the coarse-grained localization phase is: 1) positively related to the standard deviation of RSS and the distances among AP, RIS, and users; 2) negatively related to the number of RIS elements and the number of element states.
\end{remark}

\subsubsection{Localization Performance of the Fine-grained Localization Phase}

The localization loss in each cycle of the fine-grained localization phase is related to the prior probabilities. Based on~(\ref{def_approx_loss}) and~(\ref{def_d}), we have the following remark:
\begin{remark}
  If $p_{i, n}, p_{i, n'} \gg 0$, the RSS difference $|\mu_n - \mu_{n'}|$ needs to minimized to reduce the localization loss. If $p_{i, n}$ or $p_{i, n'} \approx 0$, $|\mu_n - \mu_{n'}|$ can take any value.
\end{remark}

Different from the coarse-grained localization phase where the RSS difference needs to be enlarged among all the blocks, in the fine-grained localization phase, we only need to increase the RSS differences among blocks with large prior probabilities, which implies that the RSS differences of these blocks in the optimized phase shift vector is larger than those in the coarse-grained localization phase. Therefore, the localization error in this phase is smaller than that in the coarse-grained localization phase.

The following proposition provides the localization error after multiple cycles.
\begin{proposition}
  Suppose the phase shift vector in each cycle is randomly selected, and $\mathbb{E}(s_n) = \mu_n$, where $\mathbb{E}(s_n)$ is the expectation of RSS measured by the user at the $n$-th block. When the number of cycles in the fine-grained localization phase increases, the expectation of localization error $\mathbb{E}(l_e)$ converges to zero.
  \label{pro_approx_zero}
\end{proposition}
\begin{IEEEproof}
  See Appendix~\ref{proof_approx_zero}.
\end{IEEEproof}

This proposition implies that a sufficient number of cycles and an accurate RSS model are necessary for fine-grained localization. If $\mathbb{E}(s_n) \ne \mu_n$, which means that the RSS model deviates from the actual RSS distribution and may occur in practice, the localization error cannot converge to zero.

\vspace{-3mm}
\section{Simulation Results}
\label{s_simulation}

\newcommand{\tabincell}[2]{\begin{tabular}{@{}#1@{}}#2\end{tabular}} 

\begin{table}[!t]
  \renewcommand{\arraystretch}{1.3}
  \caption{Simulation Parameters}
  \label{t_simulation}
  \centering
  \begin{tabular}{|l|l|}
    \hline
    \textbf{Parameters} & \textbf{Values}\\
    \hline
    \hline Location of the RIS's center $\bm{r}^R$ & $(0, 0, 0)$m\\
    \hline Size of the SOI $l_x\!\times\!l_y\!\times\!l_z$ & $1\!\!\times\!\!1\!\!\times\!\!1m^3$\\
    \hline Number of blocks in the SOI $N$ & $125$\\
    \hline Location of the AP $\bm{r}^A$ & $(-0.5, 0.5, 0)$m\\
    \hline Power of signal emitted by the AP $s^t$ & $30$dBm\\
    \hline Frequency of signal emitted by the AP $f_c$ & $2.4$GHz\\
    \hline Number of elements in the RIS $M$ & $25$\\
    \hline Separation between adjacent elements $d^s$ & $0.06$m\\
    \hline Number of an element's states $C$ & $4$\\
    \hline Amplitude ratio of the reflection coefficient $r$ & $1$\\
    \hline \tabincell{l}{Power gain of the AP antenna towards the \\$n$-th block $g^t_n$} & $1$\\
    \hline \tabincell{l}{Power gain of the user antenna at the $n$-th \\block towards the AP $g^r_n$} & $1$\\
    \hline \tabincell{l}{Power gain of the AP antenna towards the \\$m$-th RIS element $g^t_m$} & $1$\\
    \hline \tabincell{l}{Power gain of the user antenna at the $n$-th \\block towards the $m$-th RIS element $g^r_{m, n}$} & $1$\\
    \hline 
  \end{tabular}
\end{table}

\textcolor{black}{In this section, we present the performance of the RIS aided localization scheme in both the coarse-grained localization phase and the fine-grained localization phase. The layout of the localization scheme is shown in Fig.~\ref{f_system}, and the corresponding simulation parameters are listed in Table~\ref{t_simulation}. The RIS is on the plane $y = 0$, and the RIS center is at the origin $(0, 0, 0)$m. The SOI with size $1\times 1 \times 1$m$^3$ is divided into $125$ blocks with size $0.2 \times 0.2 \times 0.2$m$^3$, and the center of the SOI is at $(0, d, 0)$. The AP is at $(-0.5, 0.5, 0)$m, and the signal emitted by the AP has power $s^t = 30$dBm and frequency $f_c = 2.4$GHz. The RIS consists of $64$ elements, and the separation between neighboring elements is $0.06m$. Each element has $4$ states with uniform phase shift and ideal amplitude ratio, i.e., $r = 1$. We assume that the AP and the users equip omnidirectional antennas, and the power gains of the antennas $g^t_n$, $g^r_n$, $g^t_m$ and $g^r_{m, n}$ are equal to $1$. The constant $\epsilon = 1$, and parameters $Z^l$ and $Z^u$ are $2$ and $5$, respectively. The localization error in the simulation part is calculated through a Monte Carlo process, which is given by
\begin{equation}
  l_e = \dfrac{1}{IN^c}\sum^{N^c}_{n_c = 1}\sum_{i\in\mathcal{I}}||\bm{r}^e_i - \bm{r}^g_i||,
\end{equation}
where $N^C$ is the number of Monte Carlo runs. In each Monte Carlo run, the user's position are randomly chosen from all the blocks in the SOI with equal probability.}

\vspace{-3mm}
\subsection{Performance Comparison}

\textcolor{black}{To evaluate the pereformance of the proposed scheme, we provide the performance of the random phase shift scheme and some state-of-art~(SOA) RSS based schemes including the LLS~\cite{sari2018rss}, SOCP-C~\cite{chang2018rss}, SOCP-T~\cite{tomic2015rss}, DEOR~\cite{najarro2020fast} schemes. In the random phase shift scheme, the phase shift vectors are randomly set in different cycles, while the optimal decision function is adopted for location estimation.}

\begin{figure}[!t]
  \centering
  \includegraphics[width=3in]{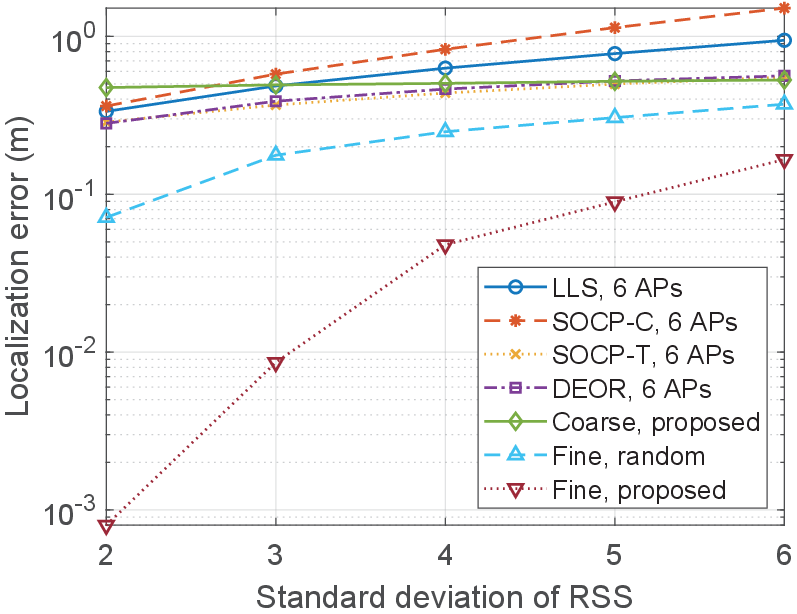}
  \caption{The localization error $l_e$ versus the standard deviation $\sigma$ for different schemes.}
  \label{f_compare}
\end{figure}

\textcolor{black}{Fig.~\ref{f_compare} presents the localization error $l_e$ versus the standard deviation $\sigma$ for different schemes when $d = 1.5$m and $I = 1$. The number of cycles in the random phase shift scheme and the proposed schemes in the fine-grained localization phase is $5$. We can observe that for all standard deviations, the proposed schemes in the fine-grained phase has the lowest localization error, which shows the superiority of the proposed scheme over other schemes\footnote{\textcolor{black}{Since the proposed scheme can provide high precision location information in 3D space, it can be used for a variety of applications such as tracking of wearable devices and indoor navigation of mobile vehicles.}}. For example, when $\sigma = 4$dB, the localization error of the proposed scheme ($l_e = 0.0479$m) in the fine-grained localization phase is more than $8$ times smaller than that of the SOCP-T scheme ($l_e = 0.4381$m). Besides, we can also observe that the proposed scheme in the coarse-grained localization phase has comparable localization accuracy with the SOA schemes, which indicates the effectiveness of customizing RSS distributions for localization accuracy improvement.}

\vspace{-3mm}
\subsection{Simulation for the coarse-grained localization phase}

\begin{figure}[!t]
  \centering
  \includegraphics[width=3in]{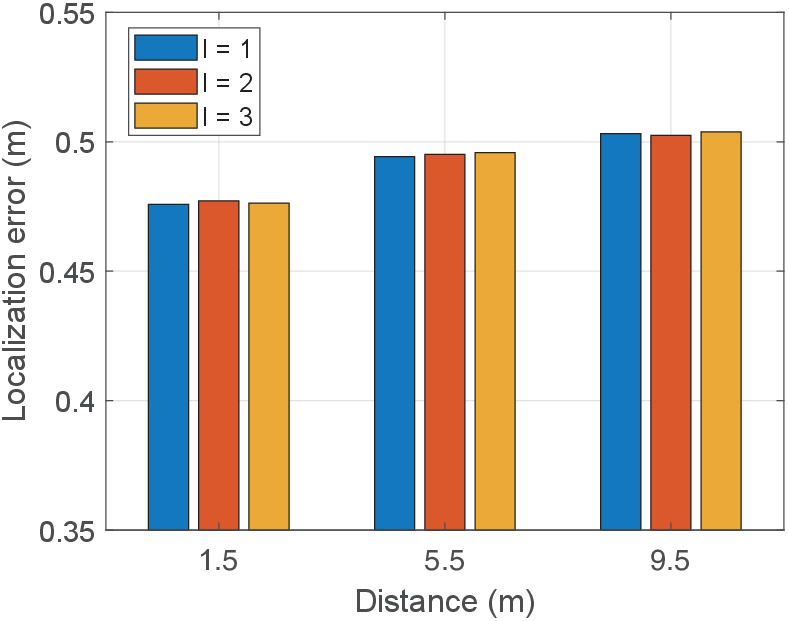}
  \caption{The localization error $l_e$ versus the distance from the RIS to the SOI $d$ in the coarse-grained localization phase.}
  \label{f_coarse_distance}
\end{figure}

Fig.~\ref{f_coarse_distance} shows the localization error $l_e$ versus the distance from the RIS to the SOI $d$ when $\sigma = 2$dB. It can be observed that the localization error $l_e$ increases with distance $d$, which is in accordance with \textbf{Remark~\ref{re_error_coarse}}. Besides, we can observe that the localization error $l_e$ remains almost unchanged when the number of users increases. This is because in the coarse-grained localization phase, the phase shift vector is fixed, and each user will not influence the localization processes of other users.

\begin{figure}[!t]
  \centering
  \includegraphics[width=3in]{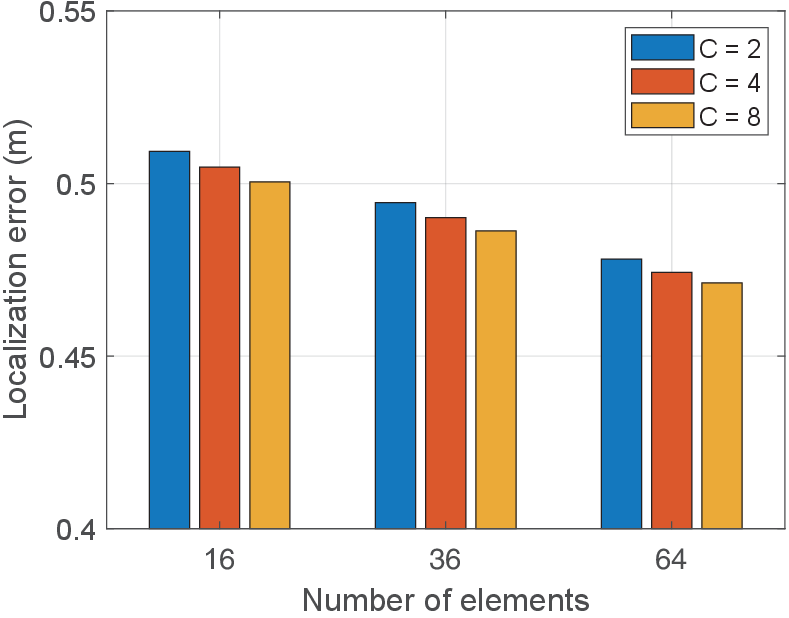}
  \caption{The localization error $l_e$ versus the number of elements $M$ in the coarse-grained localization phase.}
  \label{f_coarse_element}
\end{figure}

Fig.~\ref{f_coarse_element} depicts the localization error $l_e$ versus the number of elements $M$ when $\sigma = 2$dB and $d_S = 1.5$m. We can observe that the localization error $l_e$ decreases when the number of elements $M$ and the number of states $C$ increases, which is also consistent with \textbf{Remark~\ref{re_error_coarse}}. This indicates the trade-off between the implementation cost of RIS and the localization accuracy. Specifically, the ability of RIS to customize the RSS distribution increases with the number of elements and states, thus decreasing the localization accuracy, while the implementation cost of RIS will increase.

\vspace{-3mm}
\subsection{Simulation for the fine-grained localization phase}

\begin{figure}[!t]
  \centering
  \includegraphics[width=3in]{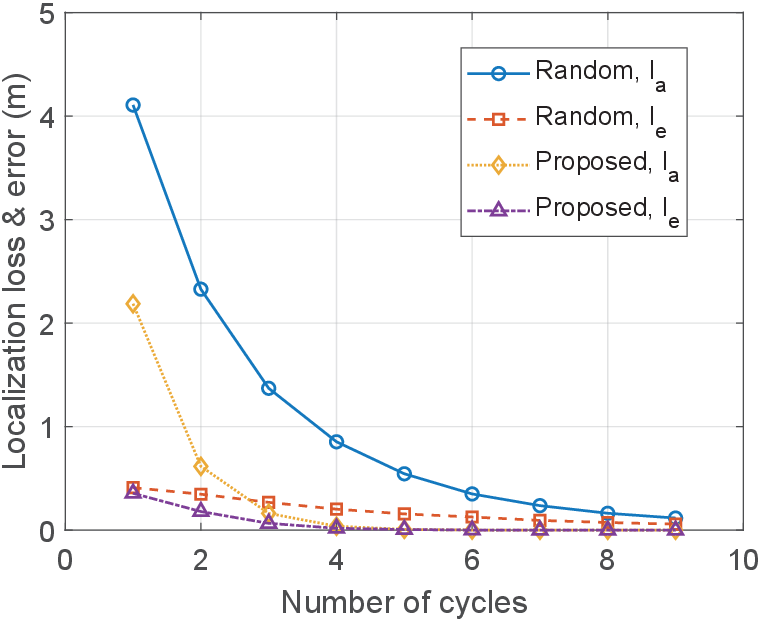}
  \caption{The localization loss $l_a$ \& error $l_e$ versus the number of cycles $K$ in the fine-grained localization phase.}
  \label{f_fine_compare}
\end{figure}

Fig.~\ref{f_fine_compare} illustrates the localization loss $l_a$ and error $l_e$ versus the number of cycles $K$ when $d = 1.5$m, $I = 1$, and $\sigma = 2$dB. We can observe that the localization errors and the localization losses of the random and proposed schemes converge to zero, which is in accordance with \textbf{Proposition~\ref{pro_approx_zero}}.
Besides, although the difference between the localization loss and the localization error is significant in the first few cycles due to the approximation when calculating the localization loss, by minimizing the localization loss in every cycle, the localization error of the proposed scheme declines faster than that of the random phase shift scheme, which shows the effectiveness of using the approximated localization loss to evaluate the localization error.

\begin{figure}[!t]
  \centering
  \includegraphics[width=2.7in]{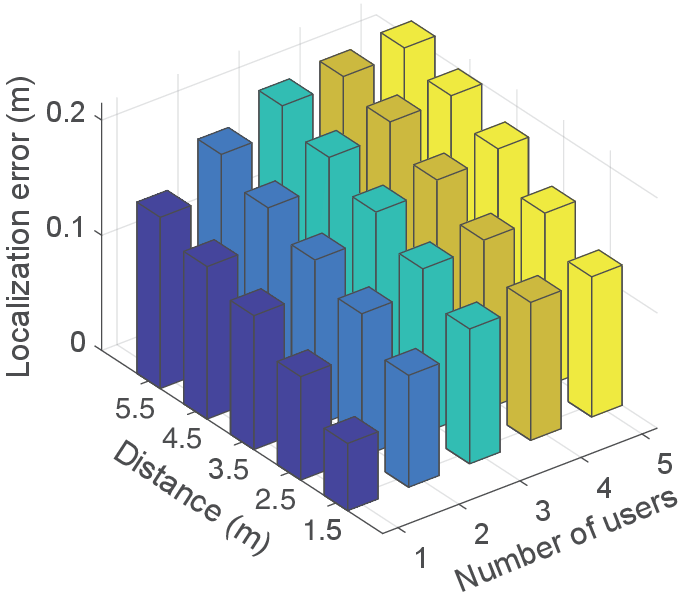}
  \caption{The localization error $l_e$ for different number of users $I$ and distance from the RIS to the SOI $d_S$.}
  \label{f_coarse_color}
\end{figure}

Fig.~\ref{f_coarse_color} depicts the localization error $l_e$ for different number of users $I$ and distance from the RIS to the SOI $d$ when $\sigma = 3$dB and the number of cycles $K = 3$. 
We can observe that the localization error $l_e$ increases with the distance $d$ and the number of users $I$.
For each user, the RSS differences between its location and the neighboring blocks need to be enlarged in the optimization problem.
When more users need to be considered simultaneously, the RSS differences among more blocks needs to be enlarged, rendering the increase of localization error.
However, the increased value of localization error decreases when the number of users increases.
This is because the neighboring blocks of users are more likely to overlap with those of others users for a larger number of users, and thus the number of these blocks does not increases linearly with the number of users.

\begin{figure}[!t]
  \centering
  \includegraphics[width=3in]{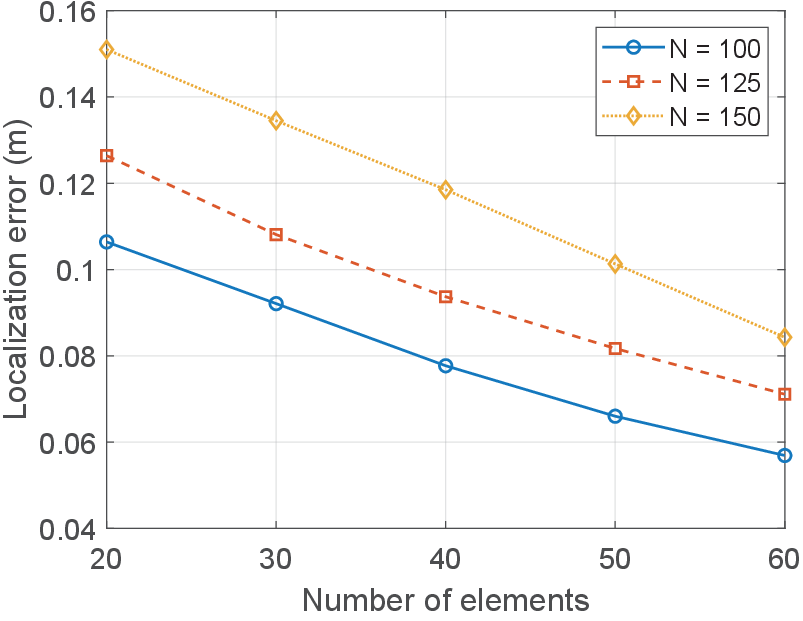}
  \caption{The localization error $l_e$ versus the number of elements $M$.}
  \label{f_fine_M}
\end{figure}

Fig.~\ref{f_fine_M} shows the localization error $l_e$ versus the number of elements $M$ when $\sigma = 3$dB and the number of cycles $K = 3$.
We can observe that the localization error $l_e$ increases with the number of blocks $N$. When $N$ increases, the RIS has to customize the radio environment in a larger range, and thus the corresponding performance will degrade.
Besides, the localization error $l_e$ decreases when the number of elements increases, which is similar to the results in the coarse-grained localization phase. 
 
\vspace{-3mm}
\subsection{Complexity}

\begin{figure}[!t]
  \centering
  \includegraphics[width=3in]{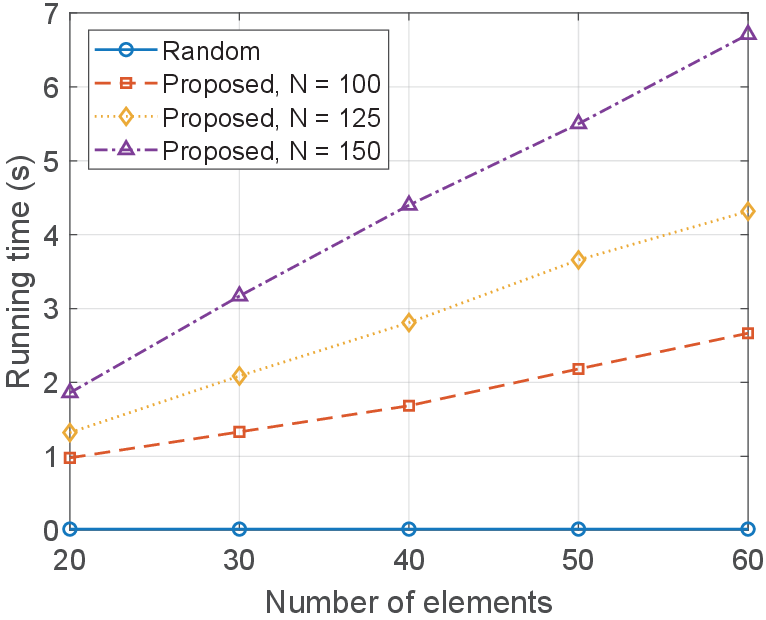}
  \caption{The simulation time $t$ versus the number of elements $M$.}
  \label{f_complexity}
\end{figure}

To test the complexity of the PSO algorithm, we use a computer with Intel Core i7-7700 CPU (3.6GHz), 16 GB RAM, and Matlab 2019b. The simulation time $t$ in each cycle versus the number of elements $M$ is shown in Fig.~\ref{f_complexity}. It can be observed that the simulation time of the proposed scheme roughly linearly increases with the number of elements $M$, and quadratically increases with the number of blocks $N$, which matches the complexity analysis in Section~\ref{ss_complexity_ro}. The running time of the random phase shift scheme remains almost unchanged when the number of elements increases. This is because we do not need to optimize the phase shifts in this scheme.

\begin{figure}[!t]
  \centering
  \includegraphics[width=3in]{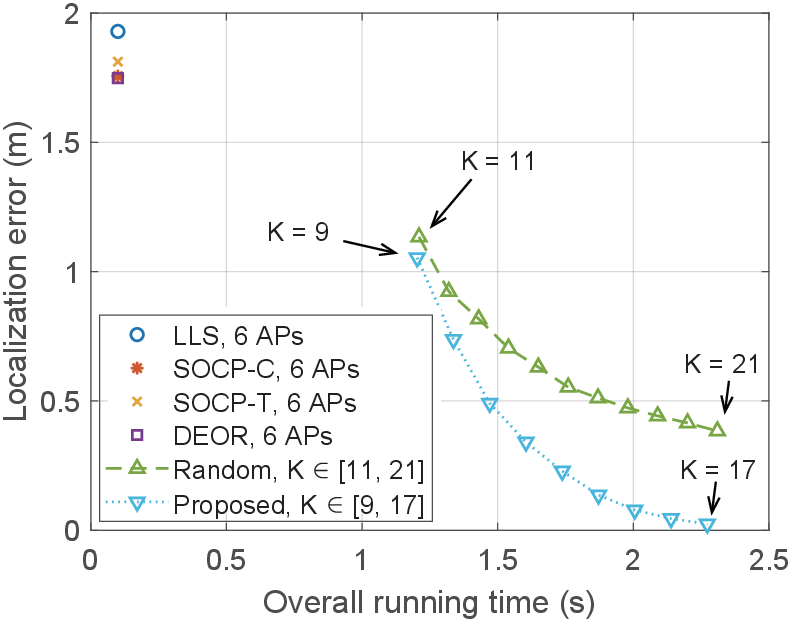}
  \caption{The localization error $l_e$ versus the overall running time $t_o$ with different localization schemes.}
  \label{f_complexity_a}
\end{figure}

\textcolor{black}{The proposed scheme is also simulated in a SOI with the size of a typical indoor environment (i.e., $l_x\times l_y\times l_z = 10\times 10\times 3$m$^3$ and $N = 37500$~\cite{qiu20163dinsar}, \cite{tlili2012accurate}). Since the algorithm complexity grows cubically with the number of blocks, the running time of the PSO algorithm becomes extremely long when $N = 37500$. To address this issue, we propose an acceleration method for the PSO algorithm. Specifically, when calculating the localization loss $l_a(\bm{c})$ in the PSO algorithm, we only consider $N_{max} \le N$ blocks with the largest probabilities, and eliminate other blocks. Let $l_a(\bm{c}, N_{max})$ denote the localization loss calculated using the corresponding $N_{max}$ blocks. Therefore, the time complexity of calculating $l_a(\bm{c}, N_{max})$ is $O(IN^2_{max})$, and the time complexity of the PSO algorithm becomes $O((Z^u + 1)I^2MN N^2_{max} + (Z^u+1-Z^l)CIN^2)$, which is much smaller than the original complexity $O((Z^u+1)(I^2MN^3))$.}

To estimate the performance of the proposed acceleration method, in Fig.~\ref{f_complexity_a}, we show the localization error $l_e$ obtained by the SOA RSS based schemes with $6$ APs, the random scheme, and the proposed scheme using the acceleration method with $N_{max} = 5$ versus the overall running time $t_o$ in the SOI with size $10\times 10\times 3$m$^3$. The overall running time of the SOA schemes is $t_o = \delta_M  + \delta_A$, where $\delta_M = 0.1$s is the duration of the RSS measurement~\cite{bahl2000radar}, \cite{Chandrasekaran2009empirical}, and $\delta_A$ is the algorithm running time. Besides, the overall running time of the random and proposed schemes is $t_o = K(\delta_A + \delta_B + \delta_M + \delta_R)$, where $K$ is the number of cycles, and $\delta_B$ and $\delta_R$ denote the durations of the broadcast and response steps, respectively. The durations $\delta_A$ and $\delta_R$ are set as $5$ms. It can be observed that the overall running times of SOA schemes are around $0.1$s, and the corresponding localization errors are in the range $[1.7, 1.9]$m. As for the random and proposed schemes, their overall running times are longer, while the localization errors can be much smaller compared with the SOA schemes. For the same overall running time, the localization error obtained by the proposed scheme is lower than that obtained by the random scheme, which shows the superiority of the proposed scheme in terms of the localization accuracy and the overall running time. Fig.~10 indicates that the SOA schemes and the proposed scheme can be used for different application scenarios with different requirements of running time and localization accuracy. Specifically, the SOA schemes are suitable for applications with high requirements of real-time performance, and the proposed scheme can be an alternative for applications requiring high localization accuracy.

\vspace{-3mm}
\section{Conclusion}
\label{s_conclusion}

In this paper, we have studied the RIS-aided multi-user wireless indoor localization using the RSS based technique. We have proposed an RIS-aided multi-user localization protocol to regulate the localization process, and formulated the optimization problem for the multi-user localization. To solve the formulated problem, we have derived the optimal decision function and designed the PSO algorithm. The effectiveness of the proposed scheme has been analysed theoretically and shown by the simulation results. According to theoretical analysis and simulation, it can be concluded that: 1) compared with traditional RSS based schemes, the localization error of the proposed scheme can be degraded by at least 3 times; 2) the localization error of the proposed scheme increases with the standard deviation of the RSS or the distance between the RIS and the SOI increases; 3) the localization error decreases when the number of RIS elements, the number of element states, or the number of cycles increases.

\appendices

\vspace{-3mm}
\section{Proof of Proposition \ref{pro_opt_decision}}
\label{proof_opt_decision}

The average localization loss can be expressed as\textcolor{black}{
\begin{align}
  l(\bm{c}, \mathcal{L}) =& \sum_{i \in \mathcal{I}} \sum_{n \in \mathcal{N}} p_{i, n} \sum_{n' \in \mathcal{N}} \gamma_{n, n'} \notag\\
  & \times \int \mathbb{P}(s_i | \bm{c}, n) \cdot \mathcal{L}(n' | \bm{c}, s_i, \{p_{i, n}\}) \cdot ds_i\notag\\
  =& \sum_{i \in \mathcal{I}} \int \bigg( \sum_{n' \in \mathcal{N}} \bigg( \sum_{n \in \mathcal{N}} p_{i, n} \gamma_{n, n'} \mathbb{P}(s_i | \bm{c}, n) \bigg)\notag\\
  &\times \mathcal{L}(n' | \bm{c}, s_i, \{p_{i, n}\}) \bigg) \cdot ds_i\notag\\
  =& \sum_{i \in \mathcal{I}}\!  \int\! \left(\sum_{n' \in \mathcal{N}}\!\! \eta_{i, n'}(s_i) \!\cdot\! \mathcal{L}(n' | \bm{c},\! s_i,\! \{p_{i, n}\}) \!\!\right) \!ds_i,
\end{align}
where $\eta_{i, n'}(s_i) = \sum_{n \in \mathcal{N}} p_{i,n} \gamma_{n, n'} \mathbb{P}(s_i | \bm{c}, n)$.} Since $p_{i, n}$, $\gamma_{n, n'}$, and $\mathbb{P}(s_i | \bm{c}, n)$ are nonnegative, we have $\eta_{i, n'}(s_i) \ge 0$. To minimize the average loss $l(\bm{c}, \mathcal{L})$, we need to choose the $n'$-th block with the minimum $\eta_{i, n'}$ as the estimated location for user $i$ with RSS $s_i$. Consequently, the decision function $\mathcal{L}$ can be expressed as
\begin{equation}
  \mathcal{L}(n' | \bm{c}, s_i, \{p_{i, n}\}) =
  \begin{cases}
    1, & s_i \in \mathcal{R}_{i, n'}\\
    0, & s_i \notin \mathcal{R}_{i, n'},
  \end{cases}
\end{equation}
where $\forall s_i \in \mathcal{R}_{i, n'}$ satisfies\textcolor{black}{
\begin{align}
  &\eta_{i, n'}(s_i) - \eta_{i, n''}(s_i) \notag\\
  =~& \sum_{n \in \mathcal{N}} \!p_{i, n} (\gamma_{n, n'} \!-\! \gamma_{n, n''}\!) \mathbb{P}(s_i | \bm{c},\! n) \!\le\! 0,\! \forall n''\! \in\! \mathcal{N}\!/\!\{n'\}.
\end{align}}

\vspace{-3mm}
\section{Proof of Proposition \ref{pro_approx_loss}}
\label{proof_approx_loss}

First, the decision region $\mathcal{R}_{i, n'}$ defined by (\ref{pro_opt_decision_c1}) is first approximated using a simpler expression. Specifically, we divide the set of all the non-negative real numbers into $N$ subsets, and the $n$-th subset is defined as
\begin{equation}
  \mathcal{R}^s_{i, n} \!\!=\! \{s_i\!\!:\! p_{i, n}\mathbb{P}(s_i | \bm{c}, n) \!\ge\! p_{i, n'}\mathbb{P}(s_i | \bm{c}, n'),\! \forall n'\!\in\!\mathcal{N}/\{n\}\!\}.  
\end{equation}
Suppose the signal-to-noise ratio~(SNR) is high and $s_i \in \mathcal{R}^s_{i, n}$, the probability $\mathbb{P}(s_i | \bm{c}, n')$ is close to $0$ if $n' \ne n$. Since $\sum_{n \in \mathcal{N}} p_{i, n} (\gamma_{n, n'} - \gamma_{n, n''}) \mathbb{P}(s_i | \bm{c}, n)$ in (\ref{pro_opt_decision_c1}) is the weighted sum of Gaussian functions with same standard deviation and different means, we can keep the term with $p_{i, n}\mathbb{P}(s_i|\bm{c}, n)$ and ignore other terms if $s_i \in \mathcal{R}^s_{i, n}$, which can be expressed as
\begin{align}
  \mathcal{R}_{i, n'} \approx \{&s_i : s_i \in \mathcal{R}^s_{i, n}, p_{i, n} (\gamma_{n,n'} - \gamma_{n,n''}) \mathbb{P} (s_i | \bm{c}, n) \le 0, \notag\\
  &\forall n'' \in \mathcal{N}/\{n'\}\}.
  \label{a2_e2}
\end{align}  
When $n = n'$, the equation $p_{i, n} (\gamma_{n,n'} - \gamma_{n,n''}) \mathbb{P} (s_i | \bm{c}, n) \le 0$ always holds because $\gamma_{n',n'} = 0$ and $\gamma_{n',n''} > 0$. If $n \ne n'$, we have $p_{i, n} (\gamma_{n,n'} - \gamma_{n,n''}) \mathbb{P} (s_i | \bm{c}, n) > 0$ when $n'' = n$. Therefore, (\ref{a2_e2}) can be expressed as
\begin{align}
  &\mathcal{R}_{i, n'} \notag\\
  \approx& \mathcal{R}^s_{i, n'} \notag\\
  =& \{s_i: p_{i, n'}\mathbb{P}(s_i | \bm{c}, n') \ge p_{i, n}\mathbb{P}(s_i | \bm{c}, n), \forall n\in\mathcal{N}/\{n'\}\}\notag\\
  =& \left\{s_i\!:\! \dfrac{\mathbb{P} (s_i | \bm{c}, n')}{\mathbb{P} (s_i | \bm{c}, n)} \ge \dfrac{p_{i, n}}{p_{i, n'}}, n\in\mathcal{N}/\{n'\}\right\}\notag\\
  =& \left\{s_i\!:\! e^{-\dfrac{(s_i - \mu_{n'})^2 - (s_i - \mu_{n})^2}{2\sigma^2}} \ge \dfrac{p_{i, n}}{p_{i, n'}}, n\in\mathcal{N}/\{n'\}\right\}\notag\\
  =& \left\{\!s_i\!:\! (s_i\! -\!\! \mu_{n'})^2\! -\! (s_i \!-\!\! \mu_{n})^2 \!\le\! 2\sigma^2 \!\ln\! \dfrac{p_{i, n'}}{p_{i, n}}\!,\! n\!\!\in\!\mathcal{N}\!/\!\{n'\}\!\!\right\}\!.
  \label{exp_approx}
\end{align}

The union bound method~\cite{proakis2007digital} can provide a tight upper bound for the region $\mathcal{R}^s_{n'}$ for high SNR. The decision region $\mathcal{R}^s_{i, n'}$ is replaced by a larger decision region $\mathcal{R}_{i, n', n}$, which can be expressed as
\begin{equation}
  \mathcal{R}_{i, n', n} \!=\! \left\{\!s_i\!:\! (s_i \!-\! \mu_{n'})^2 \!-\! (s_i \!-\! \mu_{n})^2 \!\le\! 2\sigma^2 \!\ln\! \dfrac{p_{i, n'}}{p_{i, n}}\!\right\}.\label{a2_region1}
\end{equation}
Consequently, the approximation of the localization loss can be expressed as
\begin{equation}
  l_a(\bm{c}) \!=\! \sum_{i \in \mathcal{I}}\!\sum_{n \in \mathcal{N}} \!p_{i, n}\! \sum_{n' \in \mathcal{N}} \!\gamma_{n, n'}\! \int_{s_i \in \mathcal{R}_{i, n', n}} \!\!\mathbb{P}(s_i | \bm{c}, n) \!\cdot\! ds.\label{a2_loss1}
\end{equation}

Next, we provide a simpler expression for the integral in (\ref{a2_loss1}). Specifically, according to (\ref{a2_region1}), the region $\mathcal{R}_{n', n}$ can be expressed as\small
\begin{align}
  &\mathcal{R}_{i, n', n} \notag\\
  =& \left\{ s_i: 2s_i(\mu_{n'}-\mu_n) - \mu^2_{n'} + \mu^2_n \ge - 2\sigma^2 \ln \dfrac{p_{i, n'}}{p_{i, n}}\right\}\notag\\
  =& \bigg\{ s_i: 2s_i(\mu_{n'}-\mu_n) - 2 \mu_{n}\mu_{n'} + 2\mu^2_n \ge \mu^2_{n'} + \mu^2_n - 2 \mu_{n}\mu_{n'} \notag\\
  &- 2\sigma^2 \ln \dfrac{p_{i, n'}}{p_{i, n}}\bigg\}\notag\\
  =& \left\{ s_i: (s_i - \mu_n)(\mu_{n'}-\mu_n) \ge \dfrac{1}{2}(\mu_{n'} - \mu_n)^2 - \sigma^2 \ln \dfrac{p_{i, n'}}{p_{i, n}}\right\}.
\end{align}\normalsize
Since $s_i$ follows Gaussian distribution with mean $\mu_n$ and variance $\sigma^2$, $(s_i - \mu_n)(\mu_{n'}-\mu_n)$ follows Gaussian distribution $\mathcal{N}(0, (\mu_{n'}-\mu_n)^2\sigma^2)$. Therefore, the integral in (\ref{a2_loss1}) can be expressed as
\begin{align}
  \int_{s_i \in \mathcal{R}_{i, n', n}} \!\!\mathbb{P}(s_i | \bm{c}, n)\! \cdot\! ds\! =~&\! Q\!\left(\!\dfrac{\dfrac{1}{2}(\mu_{n'} \!-\! \mu_n)^2 \!-\! \sigma^2 \ln \dfrac{p_{i, n'}}{p_{i, n}}}{\sigma|\mu_{n'}-\mu_n|}\!\right),\notag
\end{align}
where $d_{i, n, n'} = \dfrac{(\mu_{n'} - \mu_n)^2 - 2\sigma^2 \ln p_{i, n'}/p_{i, n}}{2\sigma|\mu_{n'}-\mu_n|}$.

Therefore, the localization loss can be approximated as
\begin{equation}
  l_a(\bm{c}) = \sum_{i \in \mathcal{I}}\sum_{n \in \mathcal{N}} p_{i, n} \sum_{n' \in \mathcal{N}} \gamma_{n, n'} \cdot Q(d_{i, n, n'}).
\end{equation}

\vspace{-3mm}
\section{Proof of Proposition \ref{pro_upper}}
\label{proof_upper}

The Gaussian Q function has an upper bound $1$ for any value of $d_{i, n, n'}$.As for the loss parameter $\gamma_{n, n'}$, its upper bound can be expressed as
\begin{align}
  \gamma_{n, n'} = || \bm{r}_n - \bm{r}_{n'} || < \sqrt{l^2_x + l^2_y + l^2_z}.
\end{align}

Besides, the sum of prior probabilities $\sum_{n\in\mathcal{N}} p_{i, n} = 1$. Therefore, we have
\begin{align}
  l_a(\bm{c}) =~& \sum_{i \in \mathcal{I}}\sum_{n \in \mathcal{N}} p_{i, n} \sum_{n' \in \mathcal{N}} \gamma_{n, n'} \cdot Q(d_{i, n, n'})\notag\\
  <~& I N \sqrt{l^2_x + l^2_y + l^2_z}.
\end{align}

\vspace{-3mm}
\section{Proof of Proposition \ref{pro_less}}
\label{proof_less}

\textcolor{black}{Let $\mathcal{C}^a$ denote the set of all the local minimum phase shift vectors. When using the LMVS algorithm with random phase shift vector input, the probability to generate each phase shift vector in $\mathcal{C}^a$ is equal. Therefore, we have
\begin{equation}
  \mathbb{E}(l_a(\bm{c}'')) = E^a,
\end{equation}
where $E^a$ denotes the mean localization loss of the phase shift vector in $\mathcal{C}^a$.}

In the first iteration of the global search phase, set $\mathcal{C}$ with $Z^l$ different phase shift vectors are generated using the LMVS algorithm with random phase shift vector inputs. Let $E^1$ denote the mean localization loss of phase shift vectors in $\mathcal{C}$ in the first iteration, and we have $\mathbb{E}(E^1) = E^a$ because phase shift vectors in $\mathcal{C}$ are randomly selected from $\mathcal{C}^a$. Besides, since $\bm{c}^f$ is the phase shift vector with the minimum localization loss in $\mathcal{C}$, we have $\mathbb{E}(l_a(\bm{c}^f)) < E^1 = E^a$. Using the steepest descent method, we can generate the phase shift vector $\bm{c}' = (\bm{c}^f + \zeta\bm{d})~mod~C$. If $\bm{c}' \ne \bm{c}^f$, we have $l_a(\bm{c}') < l_a(\bm{c}^f)$. Otherwise, $l_a(\bm{c}') = l_a(\bm{c}^f)$. Thus, we have
\begin{equation}
  \mathbb{E}(l_a(\bm{c}')) < \mathbb{E}(l_a(\bm{c}^f)) < \mathbb{E}(E^1) = E^a = \mathbb{E}(l_a(\bm{c}''))
\end{equation}

In the $z$-th iteration, $z-1$ phase shift vectors are added into the set $\mathcal{C}$, and the expectations of their localization loss are not greater than $E^a$. Let $E^z$ denote the mean localization loss of phase shift vectors in $\mathcal{C}$ in this iteration, and we have $\mathbb{E}(E^z) \le E^a$. Similar to proof in the first iteration, we have
\begin{equation}
  \mathbb{E}(l_a(\bm{c}')) < \mathbb{E}(l_a(\bm{c}^f)) < \mathbb{E}(E^z) \le E^a = \mathbb{E}(l_a(\bm{c}'')).
\end{equation}

\vspace{-3mm}
\section{Proof of Proposition \ref{pro_approx_zero}}
\label{proof_approx_zero}

Suppose the number of cycles in the fine-grained localization phase is $K$. The phase shift vectors are randomly selected in different cycles, which is equivalent to randomly choosing $K$ phase shift vectors at one time. Therefore, the expectation of localization error can be expressed as\small
\begin{align}
  \mathbb{E}(l_e) =~& \dfrac{\mathbb{E}(l(\bm{C}, \mathcal{L}^*))}{I} \notag\\
  =~& \dfrac{1}{I|\mathcal{C}^c|}\sum_{\bm{C} \in \mathcal{C}^c} \sum_{i \in \mathcal{I}}\sum_{\substack{n,n' \in \mathcal{N}\\ n\ne n'}} p_{i, n} \gamma_{n, n'} \int_{s_i \in \mathcal{R}_{i, n'}} \!\!\mathbb{P}(s_i | \bm{C}, n) \cdot ds,
\end{align}\normalsize
\textcolor{black}{where $\bm{C}$ is a $K \times M$ matrix, and $\mathcal{C}^c$ denotes all the possible phase shift matrices.} The $k$-th row in $\bm{C}$ denotes the phase shift vector in the $k$-th cycle. Note that the proof in Appendix~\ref{proof_approx_loss} can be extended to the circumstance where $K$ cycles are considered. According to~(\ref{exp_approx}), the decision region $\mathcal{R}_{i, n'}$ can be approximated as\small
\begin{align}
  &\mathcal{R}_{i, n'} \notag\\
  \approx& \mathcal{R}^s_{i, n'} \notag\\
  =& \left\{\bm{s}_i: |\bm{s}_i - \bm{\mu}_{n'}|^2 - |\bm{s}_i - \bm{\mu}_{n}|^2 \le 2\sigma^2 \ln \dfrac{p^0_{i, n'}}{p^0_{i, n}}, n\in\mathcal{N}/\{n'\}\right\},\notag\\
  =& \left\{\bm{s}_i: |\bm{s}_i - \bm{\mu}_{n'}|^2 \le |\bm{s}_i - \bm{\mu}_{n}|^2, n\in\mathcal{N}/\{n'\}\right\},
\end{align}\normalsize
where $\bm{s}_i = (s_{i, 1}, \cdots, s_{i, K})$ and $\bm{\mu}_n = (\mu_{n, 1}, \cdots, \mu_{n, K})$. 

Suppose that user $i$ is at the $n'$-th block, the decision region $\mathcal{R}^s_{i, n'}$ can be expressed as\small
\begin{align}
  &\mathcal{R}^s_{i, n'} \notag\\
  =& \left\{\!\bm{s}_i\!:\! |\bm{\xi}_i|^2 \le |\bm{\xi}_i + \bm{\mu}_{n'} - \bm{\mu}_{n}|^2, n\in\mathcal{N}/\{n'\}\right\},\notag\\
  =& \left\{\!\bm{s}_i\!:\! |\bm{\xi}_i|^2 \!\le\! |\bm{\xi}_i|^2 \!+\! |\bm{\mu}_{n'} \!-\! \bm{\mu}_{n}|^2 \!+\! 2\bm{\xi}_i(\bm{\mu}_{n'} \!-\! \bm{\mu}_{n}), n\!\in\!\mathcal{N}/\{n'\}\right\},\notag\\
  =& \left\{\!\bm{s}_i\!:\! |\bm{\mu}_{n'} - \bm{\mu}_{n}|^2 + 2\bm{\xi}_i(\bm{\mu}_{n'} - \bm{\mu}_{n}) \ge 0, n\in\mathcal{N}/\{n'\}\right\},
\end{align}\normalsize
where $\bm{\xi}_i = \bm{s}_i - \bm{\mu}_{n'}$. When $K \rightarrow \infty$, the first term $|\bm{\mu}_{n'} - \bm{\mu}_{n}|^2$ is greater than $0$ because the mean RSS value at different blocks can not be same for all phase shifts. Besides, the second term $2\bm{\xi}_i(\bm{\mu}_{n'} - \bm{\mu}_{n})$ converges to $0$ when $K \rightarrow \infty$ because $\bm{\xi}_i$ and $(\bm{\mu}_{n'} - \bm{\mu}_{n})$ are independent, and $\mathbb{E}(\bm{\xi}_i) = 0$. Therefore, we have $\lim_{K\rightarrow\infty} \mathbb{P}(\bm{s}_i \in \mathcal{R}^s_{i, n'}) = 1$.

Similarly, if user $i$ is not located at the $n'$-th block, we have  $\lim_{K\rightarrow\infty} \mathbb{P}(\bm{s}_i \in \mathcal{R}^s_{i, n'}) = 0$. Consequently, when $K$ increases, the expectation of localization loss converge to zero.



\ifCLASSOPTIONcaptionsoff
  \newpage
\fi



%

\vspace{20mm}
\begin{IEEEbiography}[{\includegraphics[width=1in,height=1.25in,clip,keepaspectratio]{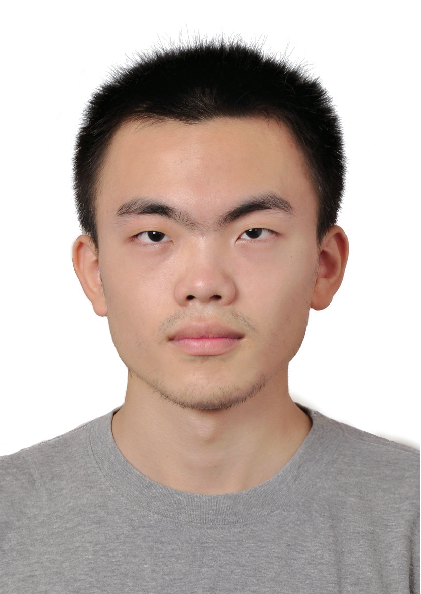}}]{Haobo Zhang} (S’19) received the B.S. degree at School of Electrical Engineering and Computer Science in Peking University in 2019, where he is currently pursuing the PhD degree in signal and information processing. His research interests include reconfigurable intelligent surfaces, wireless networks, and optimization theory.
\end{IEEEbiography}

\vspace{20mm}
\begin{IEEEbiography}[{\includegraphics[width=1in,height=1.25in,clip,keepaspectratio]{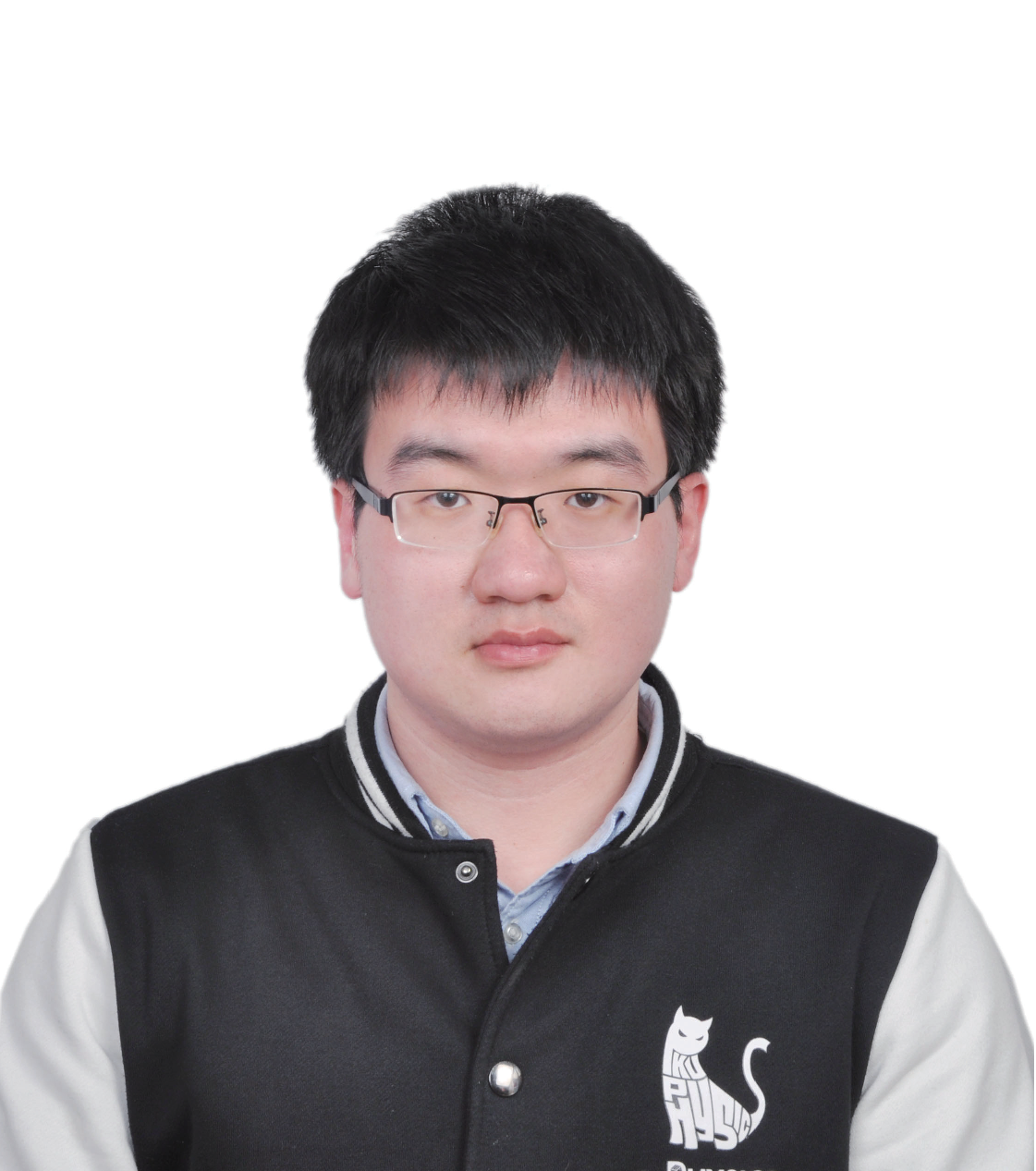}}]{Hongliang Zhang} (S’15-M’19) received the B.S. and Ph.D. degrees at the School of Electrical Engineering and Computer Science at Peking University, in 2014 and 2019, respectively. He was a Postdoctoral Fellow in the Electrical and Computer Engineering Department at the University of Houston, Texas from Jul. 2019 to Jul. 2020. Currently, he is a Postdoctoral Associate in the Department of Electrical and Computer Engineering at Princeton University, New Jersey. His current research interest includes reconfigurable intelligent surfaces, aerial access networks, and game theory. He received the best doctoral thesis award from Chinese Institute of Electronics in 2019. He is an exemplary reviewer for IEEE Transactions on Communications in 2020. He is also the recipient of 2021 IEEE Comsoc Heinrich Hertz Award for Best Communications Letters. He has served as a TPC Member for many IEEE conferences, such as Globecom, ICC, and WCNC. He is currently an Editor for IET Communications and Frontiers in Signal Processing. He also serves as a Guest Editor for IEEE IoT-J special issue on Internet of UAVs over Cellular Networks.
\end{IEEEbiography}

\begin{IEEEbiography}[{\includegraphics[width=1in,height=1.25in,clip,keepaspectratio]{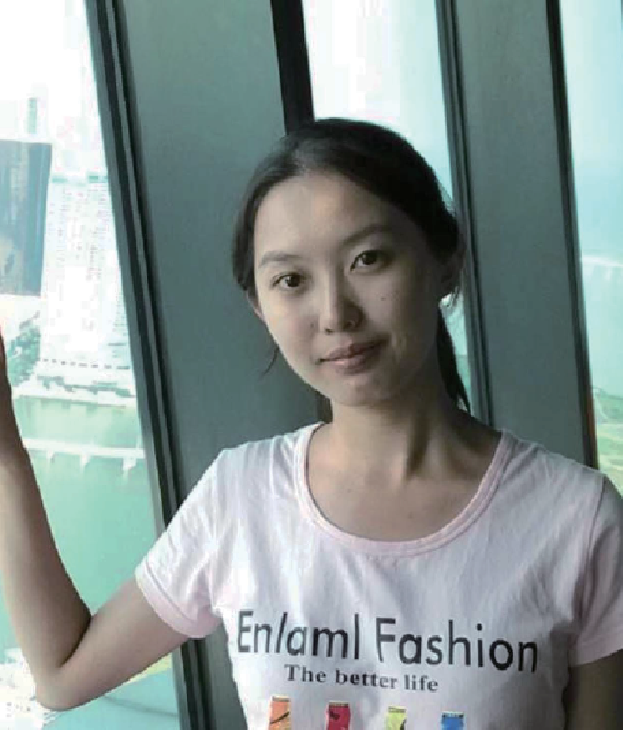}}]{Boya Di} (S'17-M'19) obtained her PhD degree from the Department of Electronics, Peking University, China, in 2019. Prior to that, she received the B.S. degree in electronic engineering from Peking University in 2014. She was a postdoc researcher at Imperial College London and is now an assistant professor at Peking University. Her current research interests include reconfigurable intelligent surfaces, multi-agent systems, edge computing, vehicular networks, and aerial access networks. One of her journal papers is currently listed as ESI highly cited papers. She serves as an associate editor for IEEE Transactions on Vehicular Technology since June 2020. She has also served as a TPC member in GlobeCom 2016, GlobeCom 2020, ICCC 2017, ICC 2016, ICC 2018, and VTC 2019.
\end{IEEEbiography}

\begin{IEEEbiography}[{\includegraphics[width=1in,height=1.25in,clip,keepaspectratio]{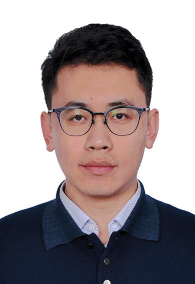}}]{Kaigui Bian} (S'05, M'11) received his Ph.D. degree in Computer Engineering from Virginia Tech in 2011, and his B.S. degree in Computer Science from Peking University, Beijing, China in 2005. He was a Visiting Young Faculty in Microsoft Research Asia in 2013. He received the best paper awards of international conferences (IEEE ICC 2015, ICCSE 2017, BIGCOM 2018) and the best student paper award of IEEE DSC 2018. He was the recipient of IEEE Communication Society Asia-Pacific Board (APB) Outstanding Young Researcher Award in 2018. He serves as an Editor for IEEE Transactions on Vehicular Technology and IEEE Access, and the organizing committee member as well as technical program committee member of many international conferences. His research interests include wireless networking and mobile computing.
\end{IEEEbiography}

\begin{IEEEbiography}[{\includegraphics[width=1in,height=1.25in,clip,keepaspectratio]{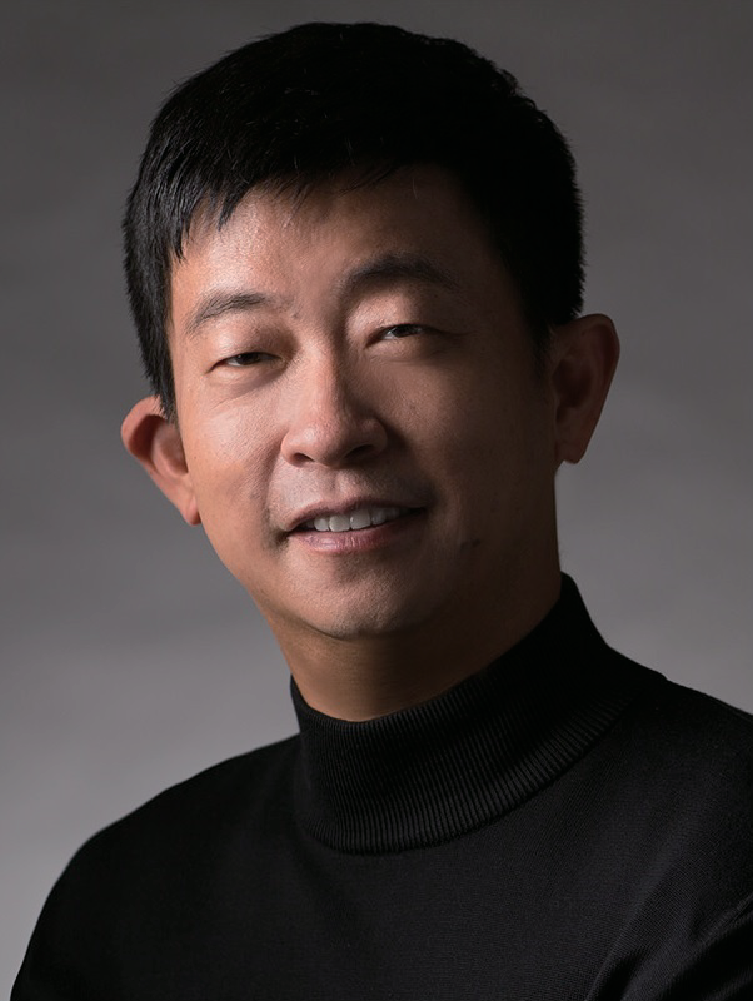}}]{Zhu Han}(S’01–M’04-SM’09-F’14) received the B.S. degree in electronic engineering from Tsinghua University, in 1997, and the M.S. and Ph.D. degrees in electrical and computer engineering from the University of Maryland, College Park, in 1999 and 2003, respectively. 

  From 2000 to 2002, he was an R\&D Engineer of JDSU, Germantown, Maryland. From 2003 to 2006, he was a Research Associate at the University of Maryland. From 2006 to 2008, he was an assistant professor at Boise State University, Idaho. Currently, he is a John and Rebecca Moores Professor in the Electrical and Computer Engineering Department as well as in the Computer Science Department at the University of Houston, Texas. His research interests include wireless resource allocation and management, wireless communications and networking, game theory, big data analysis, security, and smart grid.  Dr. Han received an NSF Career Award in 2010, the Fred W. Ellersick Prize of the IEEE Communication Society in 2011, the EURASIP Best Paper Award for the Journal on Advances in Signal Processing in 2015, IEEE Leonard G. Abraham Prize in the field of Communications Systems (best paper award in IEEE JSAC) in 2016, and several best paper awards in IEEE conferences. Dr. Han was an IEEE Communications Society Distinguished Lecturer from 2015-2018, AAAS fellow since 2019 and ACM distinguished Member since 2019. Dr. Han is 1\% highly cited researcher since 2017 according to Web of Science. Dr. Han is also the winner of 2021 IEEE Kiyo Tomiyasu Award, for outstanding early to mid-career contributions to technologies holding the promise of innovative applications, with the following citation: ``for contributions to game theory and distributed management of autonomous communication networks."
  
\end{IEEEbiography}

\begin{IEEEbiography}[{\includegraphics[width=1in,height=1.25in,clip,keepaspectratio]{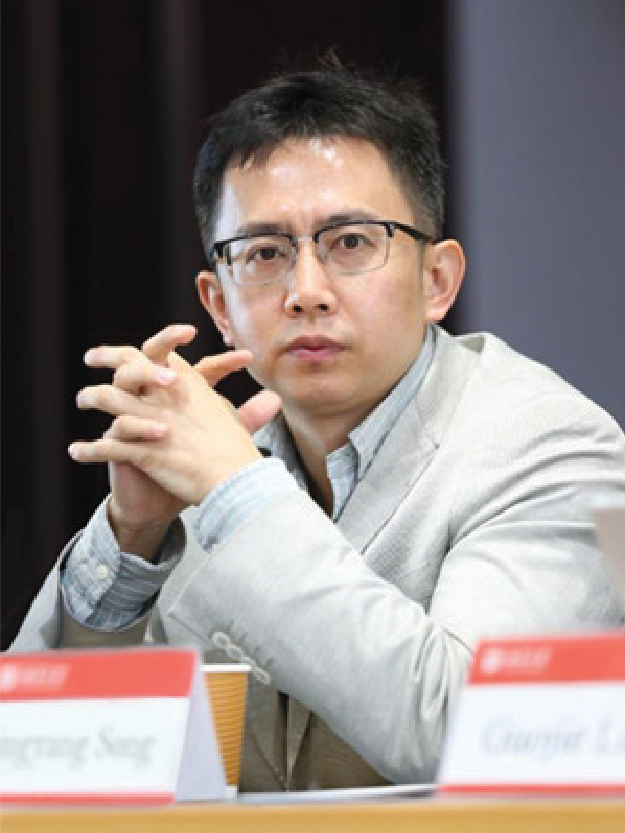}}]{Lingyang Song} (S’03–M’06-SM’11-F’19) received his PhD from the University of York, UK, in 2007, where he received the K. M. Stott Prize for excellent research. He worked as a postdoctoral research fellow at the University of Oslo, Norway, and Harvard University, until rejoining Philips Research UK in March 2008. In May 2009, he joined the School of Electronics Engineering and Computer Science, Peking University, China, as a full professor. His main research interests include cooperative and cognitive communications, physical layer security, and wireless ad hoc/sensor networks. He published extensively, wrote 6 text books, and is co-inventor of a number of patents (standard contributions). He received 9 paper awards in IEEE journal and conferences including IEEE JSAC 2016, IEEE WCNC 2012, ICC 2014, Globecom 2014, ICC 2015, etc. He is currently on the Editorial Board of IEEE Transactions on Wireless Communications and Journal of Network and Computer Applications. He served as the TPC co-chairs for the International Conference on Ubiquitous and Future Networks (ICUFN2011/2012), symposium co-chairs in the International Wireless Communications and Mobile Computing Conference (IWCMC 2009/2010), IEEE International Conference on Communication Technology (ICCT2011), and IEEE International Conference on Communications (ICC 2014, 2015). He is the recipient of 2012 IEEE Asia Pacific (AP) Young Researcher Award. Dr. Song is a fellow of IEEE, and IEEE ComSoc distinguished lecturer since 2015. 
\end{IEEEbiography}

%








\end{document}